\documentclass[pdflatex,referee,sn-mathphys-num]{sn-jnl}

\usepackage{amsmath,amssymb,amsfonts,amsthm}
\usepackage{graphicx}
\usepackage{hyperref}
\graphicspath{Figures/}
\usepackage{algorithm,algorithmicx,algpseudocode}

\begin{document}

\title{Wake-induced dynamics of soft-impact oscillators: Analysis and electronic experimental validation}

\author[a]{\fnm{Rohit} \sur{Chawla}}
\email{rohit.chawla93@gmail.com}

\author[b]{\fnm{Soumyajit} \sur{Seth}}
\email{soumyajit.seth@nmims.edu}

\author*[a]{\fnm{Vikram} \sur{Pakrashi}}
\email{vikram.pakrashi@ucd.ie}

\affil[a]{
\orgname{UCD Centre for Mechanics, Dynamical Systems and Risk Laboratory, School of Mechanical and Materials Engineering, University College Dublin},
\orgaddress{\city{Dublin}, \country{Ireland}}}

\affil[b]{
\orgname{School of Technology Management \& Engineering, SVKM's Narsee Monjee Institute of Management Studies (NMIMS)},
\orgaddress{\city{Hyderabad}, \country{India}}}

\abstract{This article presents an electronic experimental validation of discontinuity-induced bifurcations, including period-adding cascades, in a wake-induced dynamical system subjected to soft impacts. A cylindrical structure immersed in a cross-flow undergoes vortex-induced vibrations and interacts with a movable barrier during the upper and lower branches of the response. The structural motion is modelled as a harmonic oscillator, while the fluctuating lift is represented by a Van der Pol oscillator. The soft contact interaction between the structure and the movable barrier is modelled using a bilinear restoring force, resulting in a piecewise-smooth wake-oscillator system. A systematic numerical investigation reveals period-adding cascades of periodic orbits interspersed with chaotic responses as the barrier distance and fluid-flow velocity are varied. The results are experimentally validated using an inductorless electronic analogue of the piecewise-smooth wake model constructed from operational amplifiers and implemented on a printed-circuit board to improve experimental robustness. The route to instability of the periodic and aperiodic responses is further characterized using Floquet multipliers and Lyapunov exponents. A zero-time transverse discontinuity mapping is implemented to account for the evolution of perturbations in the local neighborhood of the discontinuity boundary during stability analysis. The combined numerical, experimental, and stability analyses demonstrate the role of soft impacts in generating complex nonlinear responses in the wake-induced system.}


\maketitle

\section{Introduction}\label{sec 1}

Vortex-induced vibrations (VIV) are a fundamental fluid–structure interaction phenomenon encountered in a wide range of engineering systems, particularly in offshore, marine, and energy applications. When subjected to vibro-impacts, VIV systems can exhibit rich nonlinear dynamics, particularly within the lock-in regime where the structural response synchronizes with the vortex-shedding process. In practical applications, limited clearances or obstructions and the close proximity of neighbouring structures and supports can lead to intermittent contact and vibro-impact interactions, resulting in complex periodic, aperiodic, and chaotic responses. Such large-amplitude and irregular vibrational responses are generally undesirable and can adversely affect the design, structural integrity, and operational lifetime of offshore platforms and structures \cite{Ellinas}. Examples of such coupled fluid-structure systems in the presence of neighbouring structures include tandem arrangements of marine risers \cite{xue2015practical,raiola2016wake,tian2020static}. Collision-prevention control for marine riser arrays is vital for lifetime safety and performance, since environmental loads from waves, currents, and vessel impacts can produce high stresses that may lead to failure \cite{chandrasekaran2020design}.

\noindent Examples of practical VIV systems subjected to intermittent contact or vibro-impact interactions include heat-exchanger tubes undergoing cross-flow-induced vibrations and impacting loose supports \cite{paidoussis1992cross}, floating buoyant systems connected to risers and mooring lines experiencing contact with surrounding structures \cite{virgin2009some}, and submerged structures such as large-scale trash racks in hydroelectric power plants undergoing vibration and repeated impacts \cite{xue2023nonlinear}. Related vibro-impact phenomena also arise in floating vessels subjected to sloshing-induced impact loads and in the rolling and grounding dynamics of ships \cite{ibrahim2014recent}. Such interactions are not necessarily well represented by an instantaneous rigid-impact assumption. In many practical systems, deformation of the contacting structures, supports, coatings, or other compliant components results in a finite contact duration and a force that evolves continuously during contact. Representing these interactions through a compliant, or soft-impact, contact model therefore provides a physically relevant framework for investigating the coupled influence of VIV and structural contact, including transitions between non-contacting and contacting responses and the associated nonlinear dynamical behaviour.

\noindent Low-order phenomenological models effectively capture the global dynamics of VIV systems without the computational expense of high-fidelity fluid–structure interaction simulations. In such models, the immersed structure is typically represented by a harmonic oscillator undergoing cross-flow motion, while the self-sustained fluctuating lift force responsible for VIV is described using a nonlinear wake oscillator. Phenomenological wake-oscillator models for VIV date back to the works of Bishop and Hassan \cite{bishop1964lift} and Skop and Balasubramanian \cite{balasubramanian1996nonlinear,skop1997new}, with various formulations based on the Van der Pol \cite{facchinetti2004coupling}, Rayleigh \cite{hartlen1970lift}, Rayleigh–Van der Pol oscillators \cite{krenk1999energy} and more recently \cite{ogink2010wake,qu2020single} subsequently proposed. An important aspect governing the ability of these reduced-order models to reproduce VIV is the two-way coupling between the structural and wake oscillators. Hartlen and Currie \cite{hartlen1970lift} considered velocity-based coupling, which Mureithi et al. \cite{mureithi2000bifurcation} and Plaschko \cite{plaschko2000global} later adopted. Facchinetti et al. \cite{facchinetti2004coupling} later proposed a widely adopted formulation, investigating different coupling terms and showing that acceleration coupling between the structural and wake oscillators better represented experimentally observed VIV responses. In particular, their acceleration-coupled wake-oscillator model reproduced the experimental observations of Govardhan and Williamson \cite{govardhan2000modes,govardhan2004critical} for systems with a mass ratio below the critical value of $0.54$, for which the upper response branch persists with increasing flow velocity and the synchronization range becomes unbounded. The acceleration-coupled formulation of Facchinetti et al. \cite{facchinetti2004coupling} is therefore adopted in the present study to represent the underlying VIV dynamics. To incorporate soft-impact interactions, a deformable barrier is introduced adjacent to the oscillating structure and represented through an additional stiffness and damping that become active upon contact. The resulting structural model constitutes a piecewise-smooth (PWS) bilinear oscillator coupled to the nonlinear wake oscillator, providing a low-order representation of the two-way fluid–structure interaction in the presence of soft impacts.

\noindent Despite extensive research on the VIV characteristics of cylindrical structures, the influence of intermittent contact on the lock-in regime has received comparatively limited attention, particularly from the perspective of discontinuity-induced bifurcations (DIB). In our previous work \cite{chawla2024wake}, the interaction of a structure undergoing VIV with a rigid barrier was investigated using an instantaneous impact law, where structural deformation during impact was assumed to be negligible. The rigid barrier produced DIBs such as a period-adding cascade of periodic orbits and revealed complex transitions between periodic and aperiodic responses associated with grazing and impact events. More generally, PWS systems can exhibit DIBs \cite{nusse1994border,chin1994grazing,jiang2017grazing,oestreich1996bifurcation,awrejcewicz2003bifurcation,chawla2024discontinuity} that have no direct counterpart in smooth dynamical systems, including grazing \cite{banerjee2009invisible}, period-adding cascades \cite{chawla2022stability,chawla2024discontinuity,chawla2025higherpoincare,chawla2025improved,chawla2025higher,chawla2023higher}, sticking–sliding \cite{di2002bifurcations}, and chattering \cite{budd1994chattering} bifurcations near the discontinuity boundary \cite{bernardo2008piecewise}. Grazing, in particular, occurs when a non-impacting periodic orbit tangentially encounters the contact boundary and can induce abrupt changes in the system response, including transitions to higher-period solutions \cite{chawla2024wake} and narrow-band chaotic dynamics \cite{banerjee2009invisible}.

\noindent In this article, the dynamics of a structure undergoing VIV in the presence of a pre-compressed deformable barrier are investigated using a soft-impact formulation. Unlike the instantaneous velocity change associated with a rigid impact, contact with the deformable barrier occurs over a finite duration and introduces additional stiffness and damping into the structural dynamics. The resulting system is formulated as a PWS bilinear oscillator coupled to a nonlinear wake oscillator through acceleration coupling \cite{facchinetti2004coupling,chawla2025higher}. Numerical simulations and bifurcation analysis investigate the influence of barrier stiffness and flow velocity on the system response. Particular attention is given to the transition from non-impacting to impacting motion through grazing and the subsequent evolution of periodic, multi-periodic, multi-stable, aperiodic, and chaotic responses in the VIV lock-in regime.

\noindent In addition to the numerical investigation, an analog electronic circuit is developed to experimentally validate the occurrence of DIBs due to the presence of soft impact in a VIV system, owing to its robustness compared with mechanical experimental set-ups. Conventional analog realizations of mechanical oscillators using equivalent resistor–inductor–capacitor networks can struggle to reproduce a wide range of stiffness ratios in bilinear oscillators, particularly with switching interactions at low stiffness values. In the present work, the governing differential equations are instead realized through operational amplifiers configured as analog integrators, allowing the coupled structural, wake, and soft-impact dynamics to be implemented directly. To overcome the practical difficulties of assembling a large number of interconnected analog components on a breadboard, the complete electronic system is fabricated on a dedicated printed circuit board, providing a compact and reproducible experimental platform. This electronic realization is particularly advantageous for experimentally reproducing complex DIBs, such as period-adding cascades, which can be difficult to identify systematically using conventional mechanical experimental setups. The PCB-based experiments validate the numerically predicted periodic, multi-periodic, multi-stable, and aperiodic responses and their associated bifurcations.

\noindent The stability and route to chaos of the observed dynamical responses are further investigated using Floquet theory and Lyapunov exponents. For periodic responses, the monodromy matrix characterizes the evolution of infinitesimal perturbations over one period, and the corresponding Floquet multipliers determine the stability of the periodic orbit. Since the dynamical system is nonsmooth, a zero-time transverse discontinuity mapping (ZTDM) is implemented. This ZTDM accounts for the evolution of perturbations in the local neighborhood of the discontinuity boundary during impacts and is incorporated during the construction of the monodromy matrix. Floquet analysis, however, is inherently restricted to periodic solutions and therefore cannot characterize the seemingly aperiodic responses that emerge during the period-added cascades. Lyapunov exponents are consequently evaluated to quantify the long-term evolution of infinitesimal perturbations using a ZTDM \cite{chawla2025higher} and to distinguish periodic responses from aperiodic and chaotic dynamics. The combined use of Floquet multipliers and Lyapunov exponents therefore enables determination of the stability and bifurcations of periodic solutions.

\noindent The article is structured as follows. Sec. \ref{sec 2} presents the mathematical model of the proposed PWS soft-impacting VIV system. The numerical and corresponding experimental results are presented in Secs. \ref{sec 3} and \ref{sec 4}. Sec. \ref{sec 5} discusses the stability analysis by implementing a ZTDM to account for contact dynamics. The results are summarized in Sec. \ref{sec 6}. 

\section{Mathematical model}\label{sec 2}
\begin{figure}[!]
    \centering
    \includegraphics[width=0.6\linewidth]{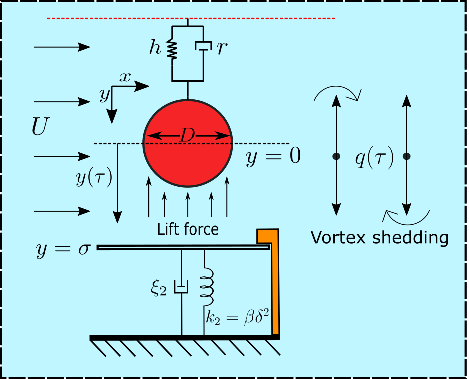}
    \caption{Schematic of a cylinder with diameter $D$ undergoing impacts with a precompressed soft barrier placed at $y = \sigma$ during onset of VIV.}
    \label{fig 1}
\end{figure}

Fig. \ref{fig 1} illustrates a cylinder of diameter $D$ immersed in a fluid medium with density $\rho$ and exhibiting VIV due to the cross-flow velocity $U$. A deformable barrier is placed near the cylinder and can softly impact the structure during large-amplitude oscillations. The structure is modelled as a linear oscillator $Y(\tau)$, and the lift coefficient $q(\tau)$ is modelled using a Van der Pol oscillator given by \cite{chawla2024wake},
\begin{align} \label{eq 1}
    m Y'' &+ hY + rY' = S,\\
    q'' &+ \Omega_f^2 q + \epsilon \Omega_f (q^2 - 1)q' = F \nonumber
\end{align}
The prime denotes the derivative with respect to dimensional time $\tau$. The oscillator mass comprises the cylinder mass and the fluid-added mass term, $m = m_s + m_f$. The oscillator damping comprises the cylinder damping and the fluid-added damping term, $r = r_s + r_f$. The oscillator stiffness is $h$, and $S$ represents the effect of the surrounding fluid on the cylinder and is proportional to the lift coefficient $q$. The lift coefficient represents a steady-state periodic force on the cylinder with vortex-shedding frequency $\Omega_f$ and nonlinear parameter $\epsilon$, and $F$ represents the effect of the cylinder's motion on the surrounding wake dynamics.

\noindent A two-way coupling low-order model, proposed by Facchinetti et al. \cite{facchinetti2004coupling,chawla2024wake}, is implemented in Eq. \eqref{eq 1}. The proposed model has two essential features. First, at the onset of VIV, the lift force $S$ exerted on the cylinder is periodic in nature, and hence a Van der Pol oscillator exhibiting a steady-state limit cycle is chosen to represent the effect of the fluid's vorticity on the cylinder during the onset of wake dynamics. Secondly, the effect of the cylinder's motion on the near-wake dynamics is given by an acceleration coupling between the cylinder and the lift coefficient, i.e., $F \propto Y''$. Facchinetti et al. \cite{facchinetti2004coupling} showed analytically that the parameters in Eq. \eqref{eq 1} can be tuned to match experimental investigations by Govardhan and Williamson \cite{govardhan2000modes,govardhan2004critical}. The proposed acceleration coupling successfully models sustained oscillations at higher cross-flow velocity $U$ when the cylinder mass is below a critical value experimentally obtained by Govardhan and Williamson \cite{govardhan2004critical}. A physical interpretation of the acceleration-coupling model is that it represents a spring-mass oscillator subjected to base excitation, with the Van der Pol oscillator externally excited by a moving structure.

\noindent Next, Eq. \eqref{eq 1} is non-dimensionalized. By rescaling $Y$ and $\tau$ as $y = Y/D$ and $t = \Omega_f \tau$, introducing oscillator's natural frequency $\Omega_s^2 = h/m$, damping coefficient $\xi_1 = r_s/2m\Omega_s$, dimensionless frequency $\delta = \Omega_s/\Omega_f$, mass ratio $\mu = m/\rho D^2$, and incorporating the soft-impact condition, Eq. \eqref{eq 1} becomes,
\begin{align} \label{eq 2}
    \ddot{y} &= \begin{cases}
        -\delta^2 y - (2 \xi_1 \delta + \gamma/\mu)\dot{y} + Mq, \qquad y < \sigma\\
        -\delta^2 y - (2 \xi_1 \delta + \gamma/\mu)\dot{y} + Mq - (\beta \delta^2 y + 2\xi_2 \dot{y}), \qquad y \geq \sigma
    \end{cases}\\
    \ddot{q} &+ q + \epsilon(q^2 - 1)\dot{q} = A\ddot{y} \nonumber
\end{align}
where the overdot is the derivative with respect to non-dimensional time $t$. Note that the following definitions were substituted in Eq. \eqref{eq 1} to obtain Eq. \eqref{eq 2}, i.e.,  $r_f = \gamma \Omega_f \rho D^2$, $\Omega_f = 2\pi S_t U/D$, $S = 1/2 \rho U^2 D C_L$, $q = 2C_L/C_{L0}$, $S/m D \Omega_f^2 M q$ and $M = C_{L0}/16\pi^2S_t^2\mu$ where $\gamma$ is the stall parameter, $S_t$ is Strouhal number, $C_L$ is the vortex lift coefficient and $C_{L0}$ is the reference lift coefficient. The fluid velocity is expressed by reduced fluid-flow velocity $U_r = 2\pi U/\Omega_s D $. The acceleration coupling is modelled as $F/D\Omega_f^2 = A \ddot{y}$. The parameter values considered in Eq. \eqref{eq 2} is given in Table \ref{tab 1}. Additionally, the cylinder interacts with the deformable barrier placed at $\sigma$ when $y \geq \sigma$ and is characterized by stiffness $\beta$ and damping ratio $\xi_2$, resulting in a Filippov system \cite{chawla2024discontinuity}. It is assumed that the soft deformable barrier placed near the cylinder does not affect or alter the fluid-flow field. Sec. \ref{sec 3} presents the effect of soft impacts on the upper branch lock-in regime during near-wake dynamics.
\begin{table}[!]
\caption{Numerical values}
\begin{tabular}{@{}lll@{}}
   \toprule
    Variable name & Symbols & Values\\
    \midrule
    Strouhal number & $S_t$ & 0.2\\
    Damping & $\xi_1$ & 0.0052\\
    Stall parameter & $\gamma$ & 0.8\\
    Reduced mass & $\mu$ & 1.19381\\
    Nonlinearity parameter & $\epsilon$ & 0.3\\
    Coupling parameter & $A$ & 12\\
    Soft-barrier damping & $\xi_2$ & 0.01\\
    \botrule
    \end{tabular}
    \label{tab 1}
\end{table}

\section{Numerical results}\label{sec 3}
The effect of soft impacts on the near-wake dynamics is explored in this section. Results were obtained using the numerical integration solver \texttt{NDSolve} in \textit{Mathematica} (v$14.3$). In every case, transient effects were discarded after a long-time integration with a relative error tolerance of $10^{-12}$.

\begin{figure}[h!]
    \centering
    \includegraphics[width=1.0\linewidth]{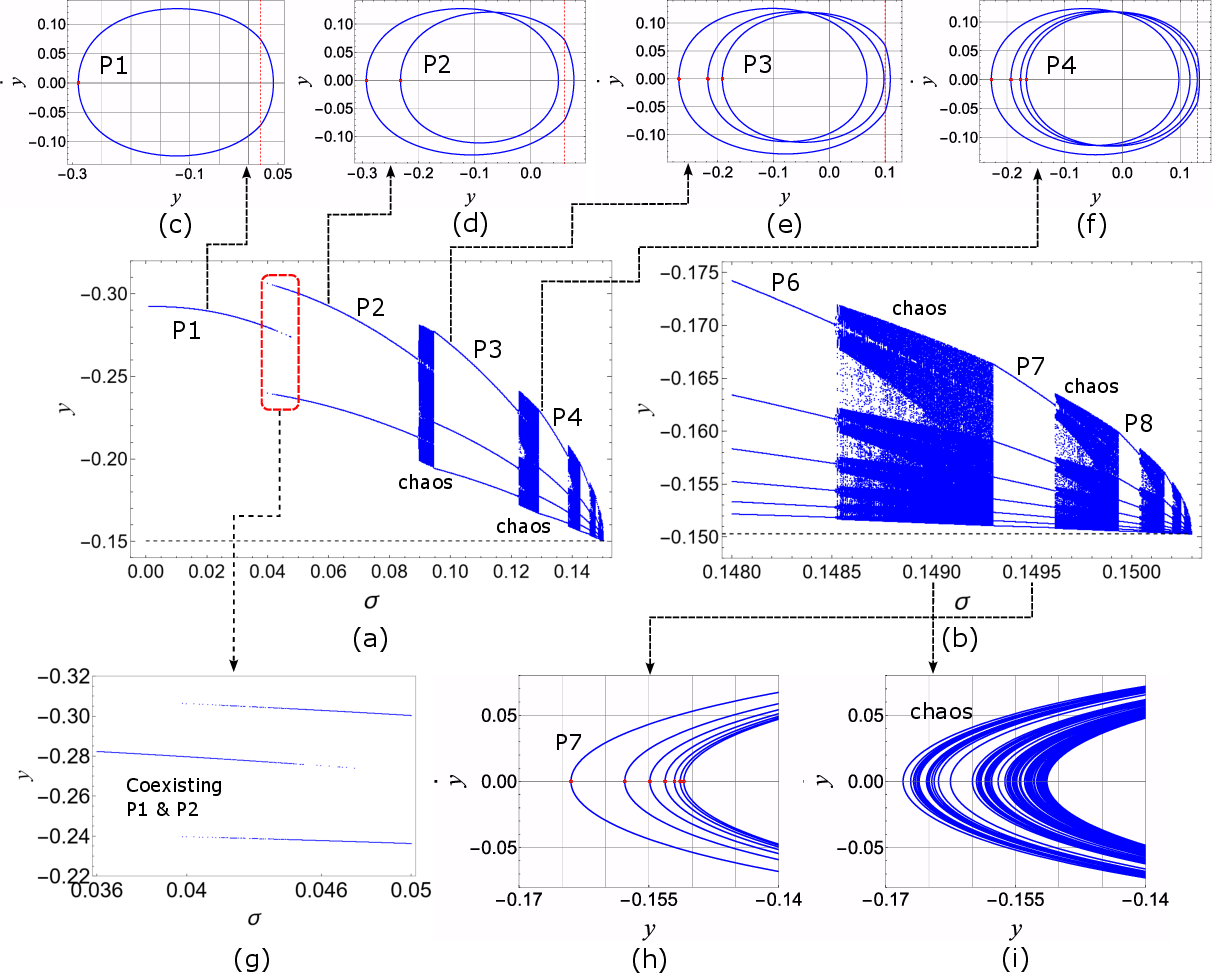}
    \caption{A period-adding cascade of orbits interspaced with aperiodic responses is presented in the bifurcation diagrams in panels (a) and (b), where the negative peak response is plotted as the barrier distance is varied over the ranges (a) $0.0 \leq \sigma \leq 0.150298$ and (b) $0.1480 \leq \sigma \leq 0.150298$. Representative periodic orbits are shown in panels (c)–(f) and (h), corresponding to P1 at $\sigma = 0.02$, P2 at $\sigma = 0.06$, P3 at $\sigma = 0.10$, P4 at $\sigma = 0.13$, and P7 at $\sigma = 0.1495$, respectively. The coexistence of P1 and P2 orbits is illustrated in panel (g) for $0.036 \leq \sigma \leq 0.05$, while a chaotic orbit is shown in panel (i) for $\sigma = 0.1490$. The red markers in the phase-space plots denote the Poincaré points associated with the negative peaks. All results are obtained for $\beta = 25.0$, $U_r = 25.0$, and $\xi_2 = 0.01$.}
    \label{fig 2}
\end{figure}

\noindent In Fig. \ref{fig 2}, the dynamics of the soft-impacting cylinder is presented as the barrier distance is varied for the stiffness ratio $\beta = 25$, reduced velocity $U_r = 25.0$ and damping $\xi_2 = 0.01$. Figure \ref{fig 2}(a) and (b) are bifurcation diagrams showing the negative peak response of the cylinder corresponding to $0.0 \leq \sigma \leq 0.150298$ and $0.1480 \leq \sigma \leq 0.150298$, respectively. A period-adding cascade of orbits is observed, interspaced by aperiodic orbits. Higher-period orbits occur near $\sigma = 0.150298$, which corresponds to the grazing value at $U_r = 25.0$. The period of an orbit is defined as the number of Poincar\'e points crossing the Poincar\'e section defined by $\dot{y} = 0$ and $\ddot{y} > 0$. These Poincar\'e points correspond to the negative peaks during the cylinder's motion. Therefore, Figs. \ref{fig 2}(c), (d), (e), (f) and (h) correspond to period 1 or simply P1, P2, P3, P4 and P7 orbits when $\sigma = 0.02$, $\sigma = 0.06$, $\sigma = 0.1$, $\sigma = 0.13$ and $\sigma = 0.1495$ respectively. Figure \ref{fig 2}(g) is a bifurcation diagram illustrating the coexistence of P1 and P2 attractors for $0.036 \leq \sigma \leq 0.05$. Within this range, as shown in (g), no aperiodic attractors are observed. A chaotic orbit is observed in Fig. \ref{fig 2}(i) for $\sigma = 0.1490$.

\begin{figure}[h!]
    \centering
    \includegraphics[width=1.0\linewidth]{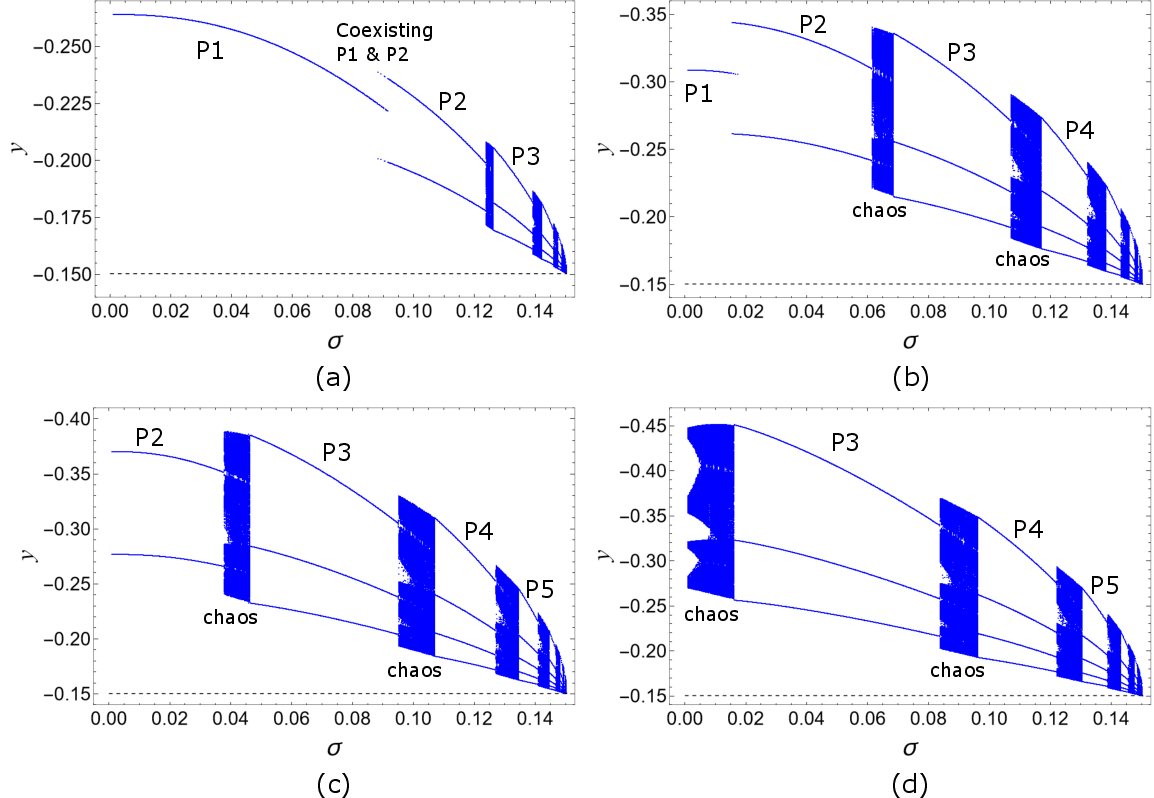}
    \caption{A period-adding cascade of orbits, interspaced with aperiodic responses, is presented in the bifurcation diagrams, where the negative peak response is plotted as the barrier distance is varied for different stiffness ratio: (a)$\beta = 10$, (b)$\beta = 50$, (c)$\beta = 100$ and (d)$\beta = 500$. All results are obtained for $U_r = 25.0$ and $\xi_2 = 0.01$.}
    \label{fig 3}
\end{figure}

\noindent The panels of Fig. \ref{fig 3} show bifurcation diagrams of the negative peak response as the barrier distance varies for different stiffness ratios, with $U_r = 25.0$ and $\xi_2 = 0.01$. Figs. \ref{fig 3}(a), (b), (c) and (d) correspond to $\beta = 10$, $50$, $100$ and $500$, respectively. The bifurcation diagrams indicate that, as the stiffness ratio increases, the period-adding cascade, interspersed with aperiodic orbits, shifts towards the lower values of $\sigma$. This eventually causes P1 and P2 orbits to disappear as the stiffness ratio increases. Additionally, whenever the P1 and P2 orbits are present at lower values of $\beta$, they coexist over a small range of $\sigma$.

\begin{figure}[h!]
    \centering
    \includegraphics[width=1.0\linewidth]{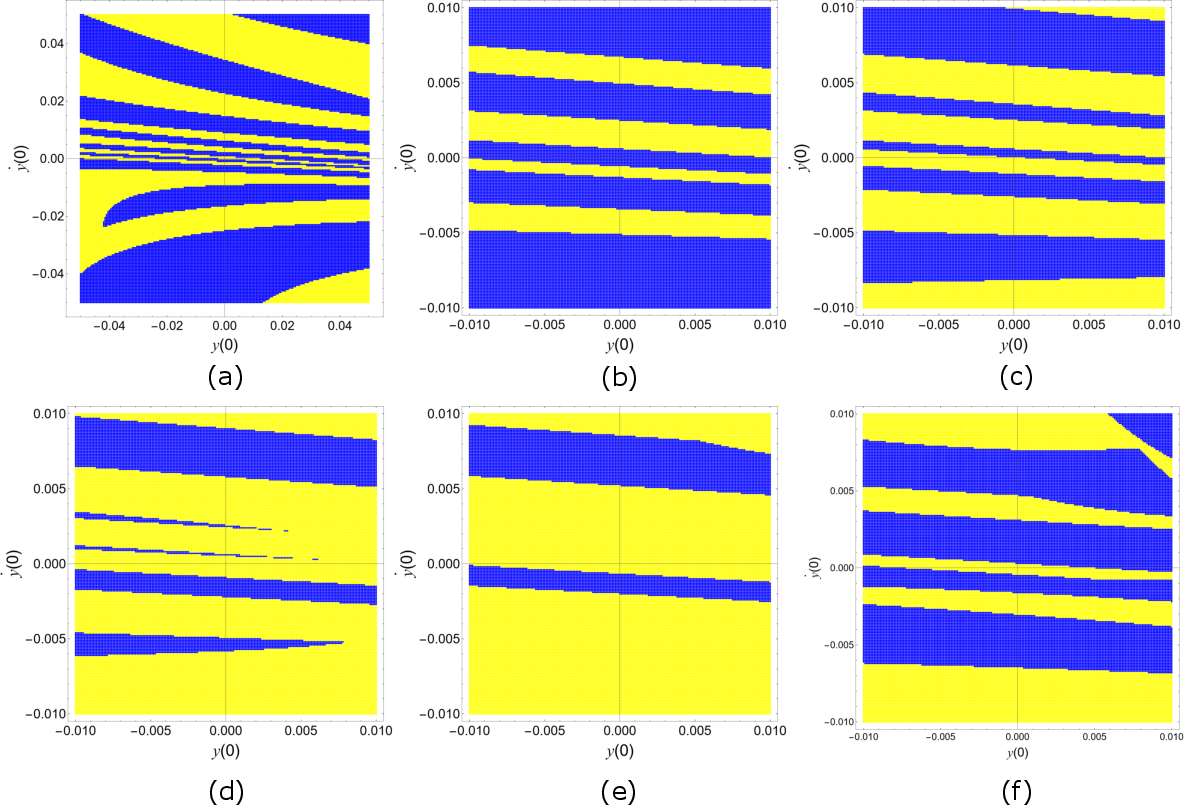}
    \caption{Basins of attraction corresponding to P1(shown in blue) and P2(shown in yellow) for varying initial conditions corresponding to (a) $\sigma = 0.09$, $\beta = 10$, (b) $\sigma = 0.068$, $\beta = 15$, (c) $\sigma = 0.055$, $\beta = 20$, (d) $\sigma = 0.045$, $\beta = 25$, (e) $\sigma = 0.030$, $\beta = 35$ and (f) $\sigma = 0.015$, $\beta = 50$. All results are obtained for $U_r = 25.0$ and $\xi_2 = 0.01$.}
    \label{fig 4}
\end{figure}

\noindent The panels in Fig. \ref{fig 4} show the basins of attraction corresponding to P1(shown in blue) and P2(shown in yellow) for varying initial conditions at different values of barrier distance $\sigma$ for $U_r = 25.0$ and $\xi_2 = 0.01$. The initial conditions of $y$ and $\dot{y}$ vary as shown in the panel while keeping $x$ and $\dot{x}$ fixed at $0.0$ and $0.01$. Figs. \ref{fig 4} (a) - (f) shows the respective P1 and P2 orbits corresponding to (a) $\sigma = 0.09$, $\beta = 10$, (b) $\sigma = 0.068$, $\beta = 15$, (c) $\sigma = 0.055$, $\beta = 20$, (d) $\sigma = 0.045$, $\beta = 25$, (e) $\sigma = 0.030$, $\beta = 35$, (f) $\sigma = 0.015$, $\beta = 50$. The results indicate that P1 and P2 orbits coexist at lower values of $\beta$ over a small range of $\sigma$.

\begin{figure}[h!]
    \centering
    \includegraphics[width=1.0\linewidth]{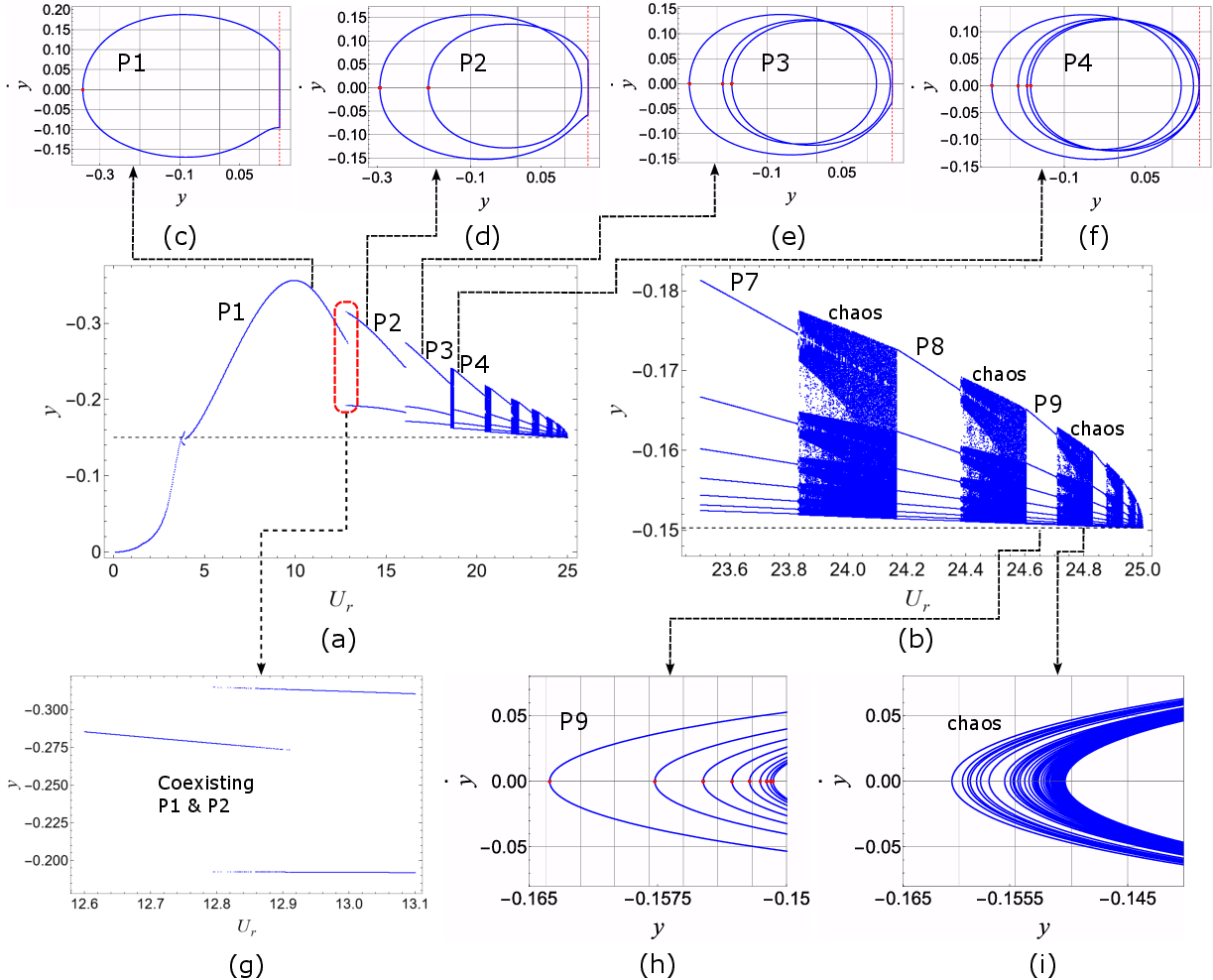}
    \caption{A period-adding cascade of orbits interspaced with aperiodic responses is presented in the bifurcation diagrams in panels (a) and (b), where the negative peak response is plotted as the reduced velocity is varied over the ranges (a) $0.0 \leq U_r \leq 25.0$ and (b) $23.5 \leq U_r \leq 25.0$. Representative periodic orbits are shown in panels (c)–(f) and (h), corresponding to P1 at $U_r = 11.0$, P2 at $U_r = 14.0$, P3 at $U_r = 17.0$, P4 at $U_r = 19.0$, and P9 at $U_r = 24.65$, respectively. The coexistence of P1 and P2 orbits is illustrated in panel (g) for $12.6 \leq U_r \leq 13.1$, while a chaotic orbit is shown in panel (i) for $U_r = 24.8$. The red markers in the phase-space plots denote the Poincaré points associated with the negative peaks. All results are obtained for $\beta = 500$, $\sigma = 0.150298$, and $\xi_2 = 0.01$.}
    \label{fig 5}
\end{figure}

\noindent In Fig. \ref{fig 5}, the dynamics of the soft-impacting cylinder is presented as the reduced velocity is varied for the stiffness ratio $\beta = 500$, barrier distance $\sigma = 0.150298$, and damping $\xi_2 = 0.01$. Figure \ref{fig 5}(a) and (b) are bifurcation diagrams showing the negative peak response of the cylinder corresponding to $0.0 \leq U_r \leq 25.0$ and $23.5 \leq U_r \leq 25.0$, respectively. A period-adding cascade of orbits is observed interspaced by aperiodic orbits. Higher-period orbits occur near $U_r = 25.0$, as this corresponds to the grazing value at $\sigma = 0.150298$. Figs. \ref{fig 5}(c), (d), (e), (f) and (h) correspond to P1, P2, P3, P4 and P9 orbits when $U_r = 11.0$, $U_r = 14.0$, $U_r = 17.0$, $U_r = 19.0$ and $U_r = 24.65$, respectively. Figure \ref{fig 5}(g) is a bifurcation diagram illustrating the coexistence of P1 and P2 attractors for $12.6 \leq U_r \leq 13.1$. Within this range, as shown in (g), no aperiodic attractors are observed. A chaotic orbit is observed in Fig. \ref{fig 5}(i) for $U_r = 24.8$.

\begin{figure}[h!]
    \centering
    \includegraphics[width=1.0\linewidth]{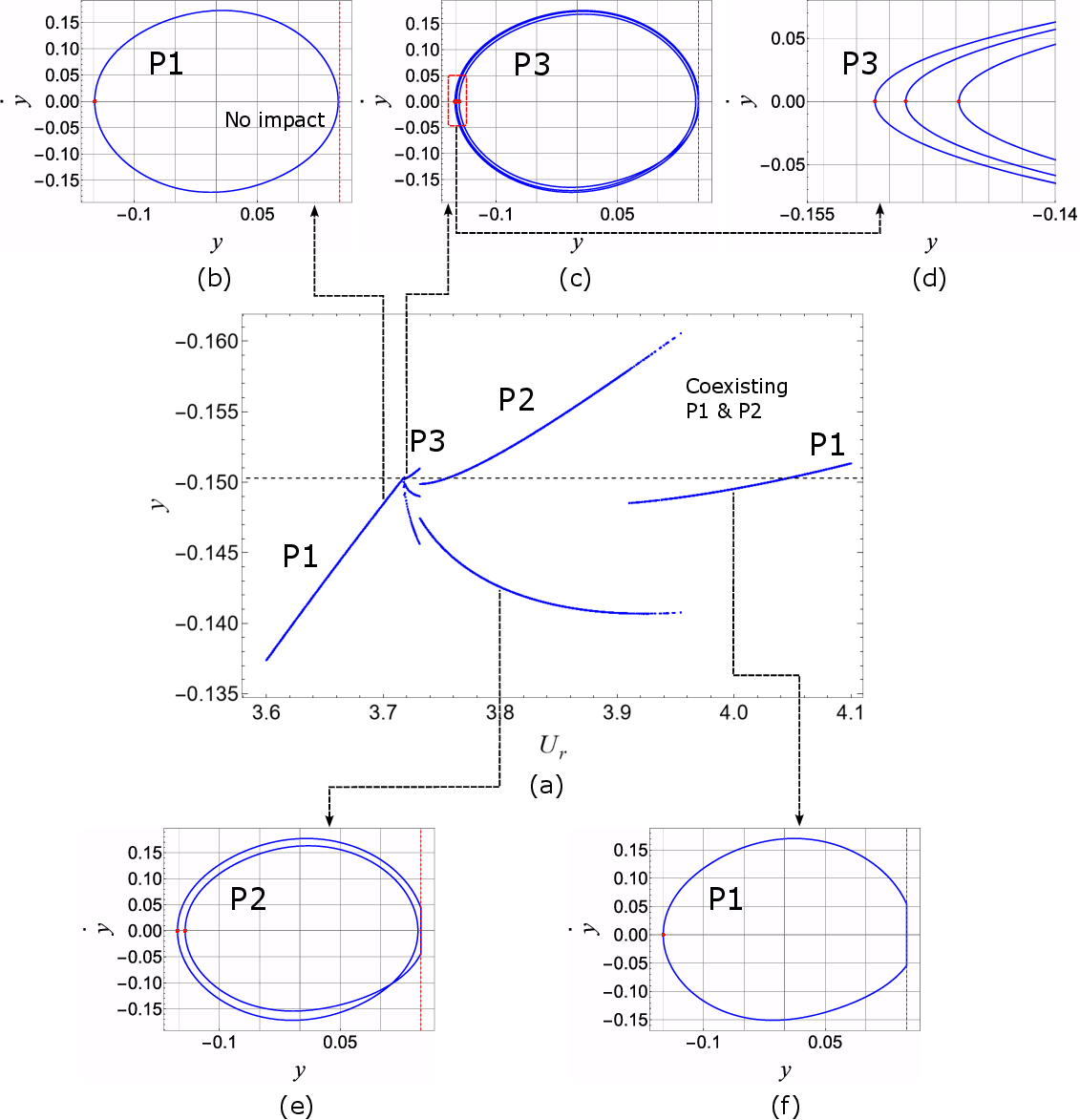}
    \caption{A bifurcation diagram is shown in panel (a) where the negative peak response is plotted as the reduced velocity is varied between $3.6 \leq U_r \leq 4.1$. Representative periodic orbits are shown in panels (b)–(f), corresponding to P1(no impact) at $U_r = 3.7$, P3 at $U_r = 3.72$ and its corresponding zoomed panel in (d), P2 at $U_r = 3.8$, and P1 at $U_r = 4.0$, respectively. The coexistence of P1 and P2 orbits can be observed in panel (a). The red markers in the phase-space plots denote the Poincaré points associated with the negative peaks. All results are obtained for $\beta = 500.0$, $\sigma = 0.150298$, and $\xi_2 = 0.01$.}
    \label{fig 6}
\end{figure}

\noindent Fig. \ref{fig 6} shows the dynamics of the soft-impacting cylinder near the initial-upper-branch transition as the reduced velocity varies for stiffness ratio $\beta = 500$, barrier distance $\sigma = 0.150298$, and damping $\xi_2 = 0.01$. Figure \ref{fig 6}(a) is a bifurcation diagram showing negative peak response of the cylinder corresponding to $3.6 \leq U_r \leq 4.1$. Initially, P1 orbits are observed when there is no impact. As $U_r$ is increased, a reverse period-adding cascade is observed, resulting in P3, P2, and P1 orbits. These are shown in Figs. \ref{fig 6}(b) - (f) correspond to P1 (no impact) at $U_r = 3.7$, P3 at $U_r = 3.72$ and its corresponding zoomed panel in (d), P2 at $U_r = 3.8$, and P1 at $U_r = 4.0$, respectively. The coexistence of P1 and P2 orbits can be observed in the bifurcation diagram of Fig. \ref{fig 6}(a).

\noindent The following section experimentally validates a period-adding cascade of orbits using an analog electronic circuit that reproduces the DIBs occurring in the soft-impacting Filippov system described above. 

\section{Experimental results}\label{sec 4}
The DIBs observed, including the period-adding cascade, in Sec. \ref{sec 3} are experimentally validated using an analog electronic circuit. The electronic circuit comprises operational amplifiers (op-amps) configured as integrators. Two op-amps integrate two first-order ODEs to model each oscillator described in Eqs. \ref{eq 2}. Thus, using the standard op-amp integrator expressions, the soft-impacting system can be implemented electronically as follows:
\begin{align}\label{eq 3}
    V^y_1 &= \frac{-1}{R C_1} \int (-\frac{R_2}{R_1}V^y_2)dt \\
    V^y_2 &= \begin{cases}
        \frac{-1}{R C_2} \int (\frac{R}{R_1^y}V^y_1 + \frac{R}{R_2^y}V^y_2 - \frac{R}{R_3^y}V^x_1)dt \qquad V_1^y < V_{\text{ref}}\\
        \frac{-1}{R C_2} \int (\frac{R}{R_1^y}V^y_1 + \frac{R}{R_2^y}V^y_2 - \frac{R}{R_3^y}V^x_1 + \frac{R}{R_4^y}V^y_1 + \frac{R}{R_5^y}V^y_2)dt \qquad V_1^y \geq V_{\text{ref}}
    \end{cases} \nonumber \\
    V^x_1 &= \frac{-1}{R C_3} \int (-\frac{R_4}{R_3}V^x_2)dt \nonumber \\
    V^x_2 &= \begin{cases}
        \frac{-1}{R C_4} \int (\frac{R}{R_1^x}V^x_1 + \frac{R}{R_2^x}(V^x_1)^2 V^x_2 - \frac{R}{R_3^x}V^x_2 + \frac{R}{R_4^x}V^y_1 + \frac{R}{R_5^x}V^y_2 - \frac{R}{R_6^x}V^x_1)dt \qquad V_1^y < V_{\text{ref}} \\
        \frac{-1}{R C_4} \int (\frac{R}{R_1^x}V^x_1 + \frac{R}{R_2^x}(V^x_1)^2 V^x_2 - \frac{R}{R_3^x}V^x_2 + \frac{R}{R_4^x}V^y_1 + \frac{R}{R_5^x}V^y_2 - \frac{R}{R_6^x}V^x_1 + \frac{R}{R_7^x}V^y_1 + \frac{R}{R_8^x}V^y_2)dt \qquad V_1^y \geq V_{\text{ref}} \nonumber
    \end{cases}
\end{align}
where the voltages $V_1^y$ and $V_2^y$ represent the structure displacement and velocity, and $V_1^x$, and $V_2^x$ represent the wake variable and its derivative. The time scale of the electrical representation of the dynamical system is controlled by the resistance $R = 10\text{K}\Omega$, and $C_1 = C_2 = C_3 = C_4 = C = 10\text{nF}$, resulting in a time constant of $RC = 10^{-4}$s. This leads to a circuit frequency of $f = \frac{1}{2\pi RC} \approx 1.591$kHz. Therefore, one time unit in Eq. \ref{eq 2} corresponds to a circuit time unit of $10^{-4}$s resulting in a time-series voltage with a frequency range of $1.59$kHz. The resistors $R_1 = R_2 = R_3 = R_4 = 10\text{K}\Omega$ ensure a unity gain to invert the sign before integrating $V_2^y$ and $V_2^x$. The following resistors ensure that the circuit equations model the parameters given in Table \ref{tab 1},
\begin{align}\label{eq 4}
    R_1^y &= \frac{R}{\delta^2} = 250\text{K}\Omega; \quad R_2^y = \frac{R}{(2\xi_1\delta + \frac{\gamma}{\mu})} = 14.87\text{K}\Omega; \quad R_3^y = \frac{R}{M} = 238.76\text{K}\Omega;\\
    R_4^y &= \frac{R}{\beta\delta^2} = 10\text{K}\Omega; \quad R_5^y = \frac{R}{2\xi_2} = 500\text{K}\Omega;\nonumber\\
    R_1^x &= \frac{R}{1} = 10\text{K}\Omega; \quad R_2^x = \frac{R}{100\epsilon} = 333.33\Omega; \quad R_3^x = \frac{R}{\epsilon} = 33.33\text{K}\Omega; \nonumber \\
    R_4^x &= \frac{R}{A\delta^2} = 20.833\text{K}\Omega; \quad R_5^x = \frac{R}{A(2\xi_1\delta + \frac{\gamma}{\mu})} = 1.239\text{K}\Omega; \quad R_6^x = \frac{R}{AM} = 19.89\text{K}\Omega; \nonumber \\
    R_7^x &= \frac{R}{A\beta\delta^2} = 833.33\Omega; \quad R_8^x = \frac{R}{2A\xi_2} = 41.66\text{K}\Omega \nonumber
\end{align}
\noindent Note that $R_2^x$ represents the cubic nonlinear term $x^2\dot{x}$. The analog multiplication is performed using two cascaded AD633 multiplier ICs, each introducing a scaling factor of $1/10$. Consequently, the resulting output is scaled by a factor of $1/100$. The factor of $10^2$ incorporated into the expression for $R_2^x$ in Eq. \ref{eq 4} compensates for this scaling and ensures the correct representation of the nonlinear term in Eq. \ref{eq 2}.

\begin{figure}[h!]
    \centering
    \includegraphics[width=1.0\linewidth]{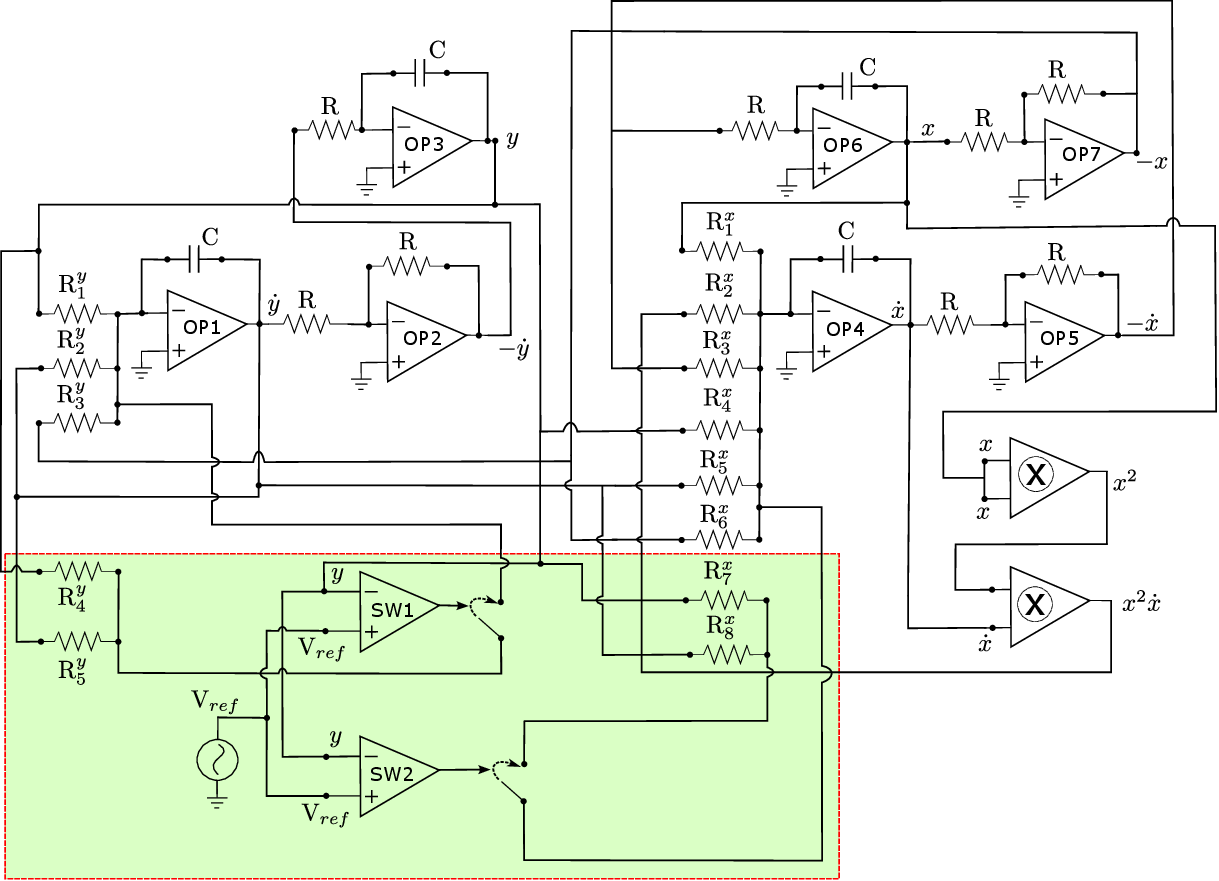}
    \caption{Schematic circuit diagram that represents the PWS soft-impacting oscillator. The shaded green region highlights the switching circuit which gets activated when $y$ or its equivalent $V_1^y \geq V_\mathrm{ref}$.}
    \label{fig 7}
\end{figure}

\noindent Figure \ref{fig 7} show the circuit diagram of the PWS soft-impacting bilinear oscillator. An impact event is controlled by the reference voltage $V_{\text{ref}}$, which sets the location of the discontinuity boundary $\sigma$. A comparator integrated circuit (IC LM311p) monitors the cylinder voltage $V_1^y$ and activates when $V_1^y \geq V_{\text{ref}}$. The comparator IC then sends an ON signal to a demultiplexer IC (IC CD4051BE), which acts as a switch, thereby connecting the additional circuit components that represent the soft impact/contact dynamics \cite{chawla2025higher}. 

\begin{figure}[h!]
    \centering
    \includegraphics[width=1.0\linewidth]{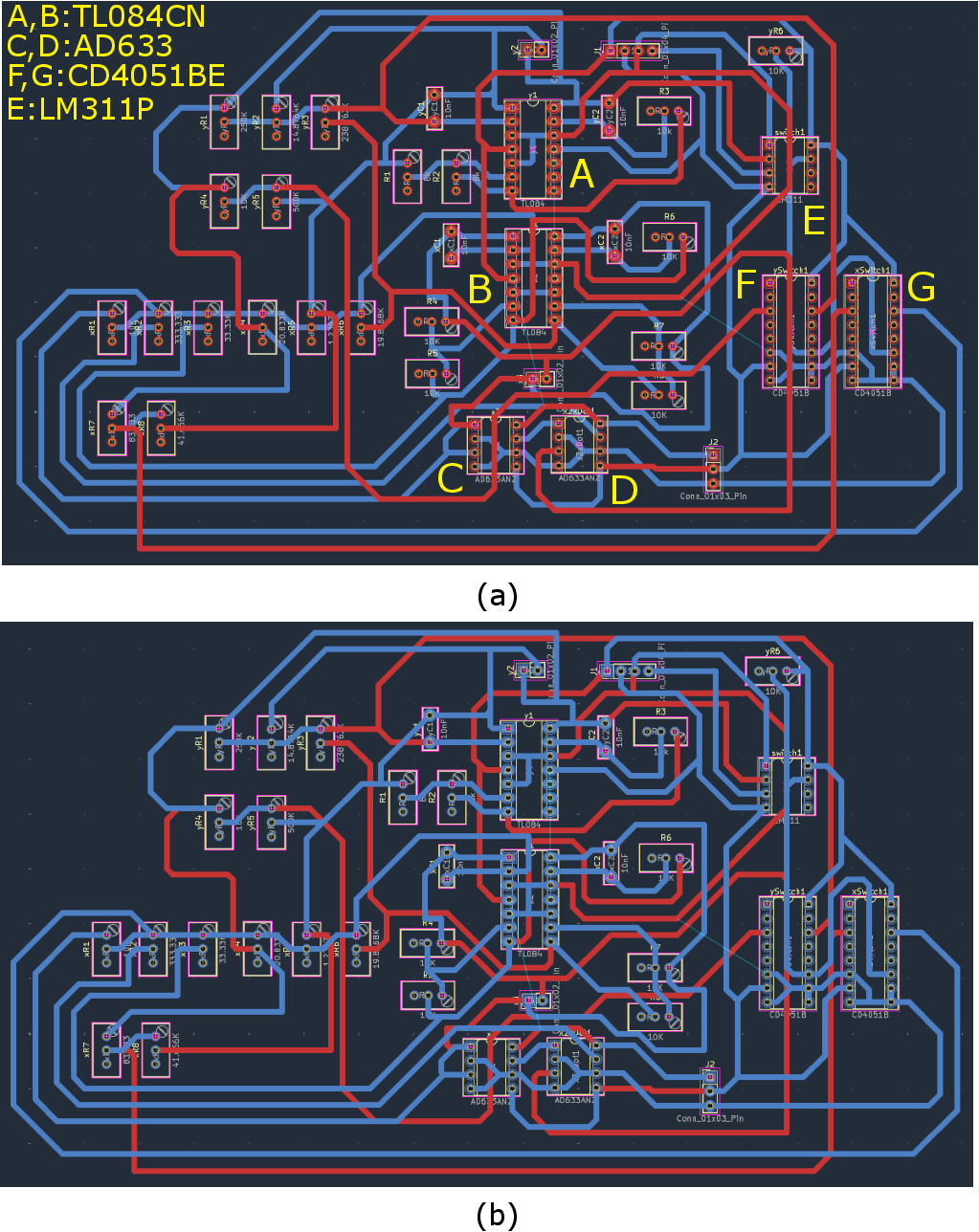}
    \caption{The PCB design showing (a) top layer and (b) bottom layer implemented using KiCAD along with the respective IC positions.}
    \label{fig 8}
\end{figure}

\begin{figure}[h!]
    \centering
    \includegraphics[width=1.0\linewidth]{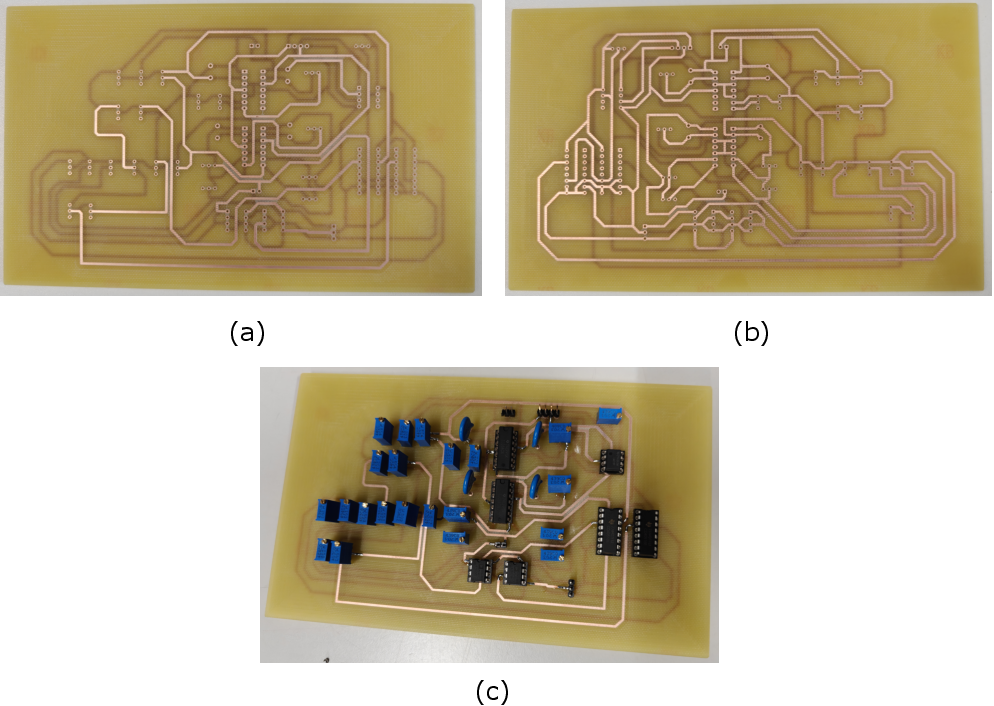}
    \caption{The top and bottom layers of the fabricated PCB is shown in (a) and (b). The fabricated PCB with components soldered in place is shown in (c).}
    \label{fig 9}
\end{figure}

\noindent To minimize noise from loose connections, the circuit is built on a printed-circuit board (PCB) designed in KiCAD. Figs. \ref {fig 8}(a) and (b) show the top and bottom PCB layout along with the positions of implemented ICs. The top and bottom layers of the fabricated PCB is shown in Figs. \ref{fig 9}(a) and (b). The fabricated PCB with components soldered in place is shown in \ref{fig 9}(c). 

\begin{figure}[h!]
    \centering
    \includegraphics[width=1.0\linewidth]{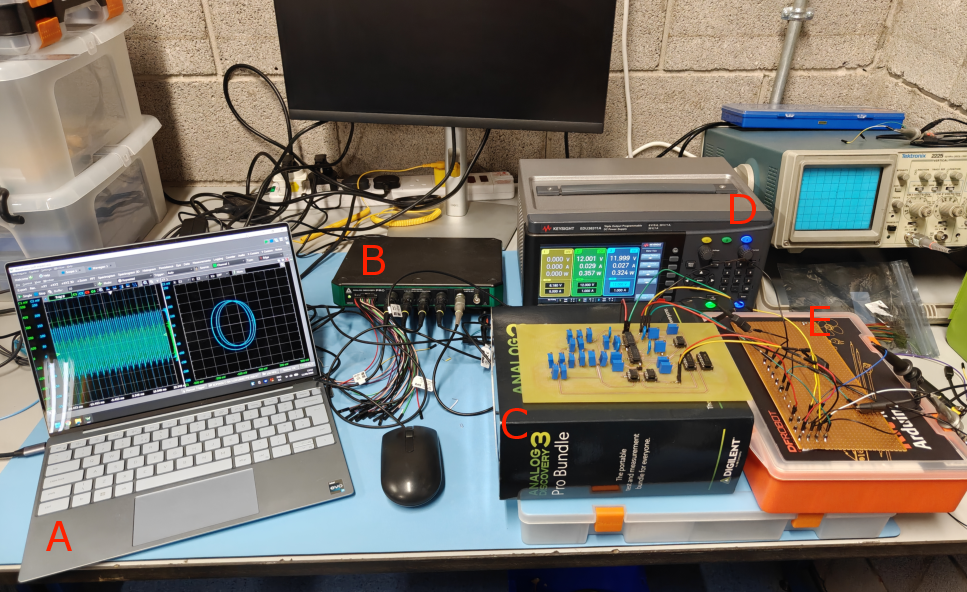}
    \caption{The experimental set-up comprises of a laptop (A), a Digilent 4-channel DAQ Analog Discovery Pro ADP3450 (B), the fabricated pcb (C), a $\pm12.0$V power source (D) and a veroboard for grounding plane with power connectors (E).}
    \label{fig 10}
\end{figure}

\noindent The experimental setup is shown in Fig. \ref{fig 10}. The setup comprises a Digilent four-channel data acquisition (DAQ) device (Analog Discovery Pro ADP3450) that simultaneously measures $V_1^y$, $V_2^y$, and $V_{\mathrm{ref}}$. Digilent WaveForms and MATLAB's Data Acquisition Toolbox are used for acquiring and recording the voltage signals. The experimental setup is placed on an electrostatic discharge (ESD) mat to minimise electrostatic charge accumulation. Appropriate signal grounding is maintained between the PCB, power supply, and DAQ to minimise electrical interference, as the PWS oscillator is sensitive to electrical noise, particularly near the discontinuity boundary.

\begin{figure}[h!]
    \centering
    \includegraphics[width=1.0\linewidth]{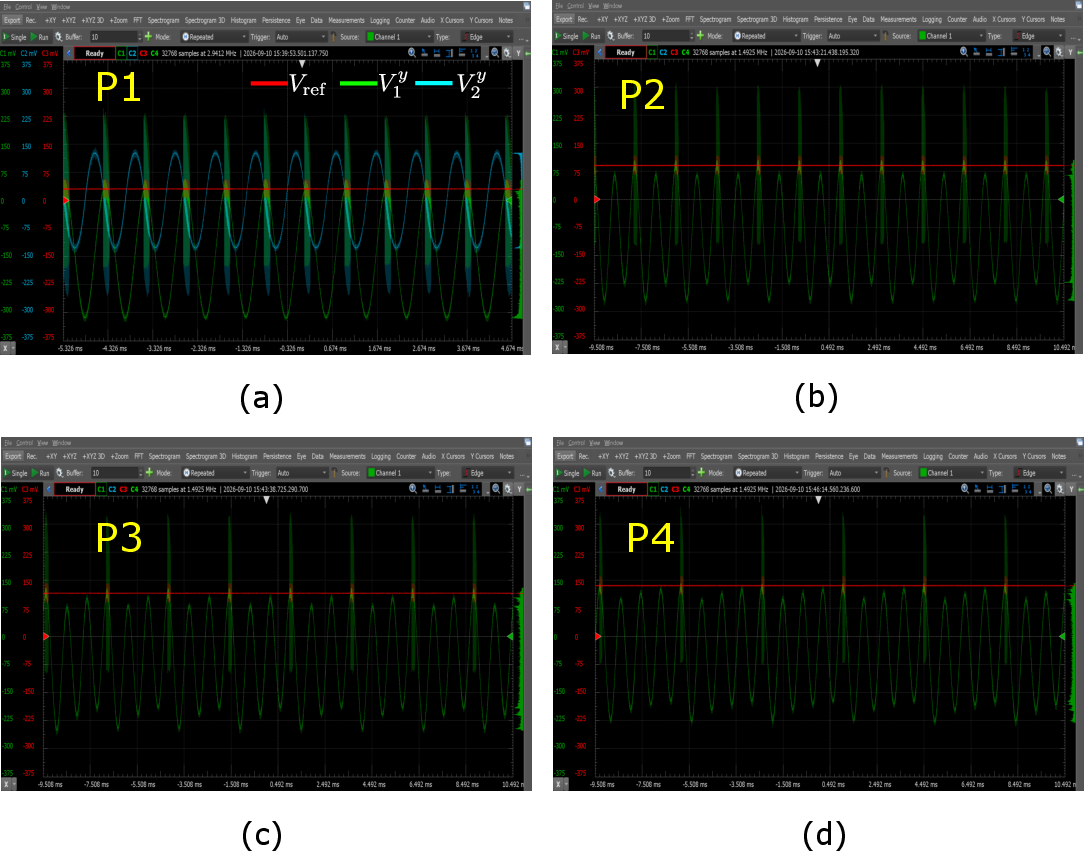}
    \caption{Experimental time series response of $V_1^y$ corresponding to (a) P1 for $V_{\mathrm{ref}} = 30$mV, (b) P2 for $V_{\mathrm{ref}} = 90$mV, (c) P3 for $V_{\mathrm{ref}} = 115$mV and (d) P4 for $V_{\mathrm{ref}} = 135$mV. The reference voltage $V_{\mathrm{ref}}$ that represents the location of the barrier $\sigma$ is shown in red.}
    \label{fig 11}
\end{figure}

\noindent The panels in Fig. \ref{fig 11} show the experimentally obtained time series response of $V_1^y$ corresponding to cylinder displacement $y$ for different reference voltages, $V_{\textrm{ref}}$. Fig. \ref{fig 11}(a) shows a P1 orbit when $V_{\text{ref}} = 30$mV. The green, blue and red plots correspond to $V_1^y$, $V_2^y$ and barrier placed at $V_{\textrm{ref}} = 30$mV. The plot highlights when $V_1^y$ becomes greater than $V_{\textrm{ref}}$ and undergoes a soft-impact. Similarly, in Figs. \ref{fig 11}(b)-(d), as the barrier location, characterised by $V_{\textrm{ref}}$, increases, P2, P3 and P4 orbits are observed corresponding to $V_{\textrm{ref}}$ of $90$, $115$ and $135$mV. 

\begin{figure}[h!]
    \centering
    \includegraphics[width=0.9\linewidth]{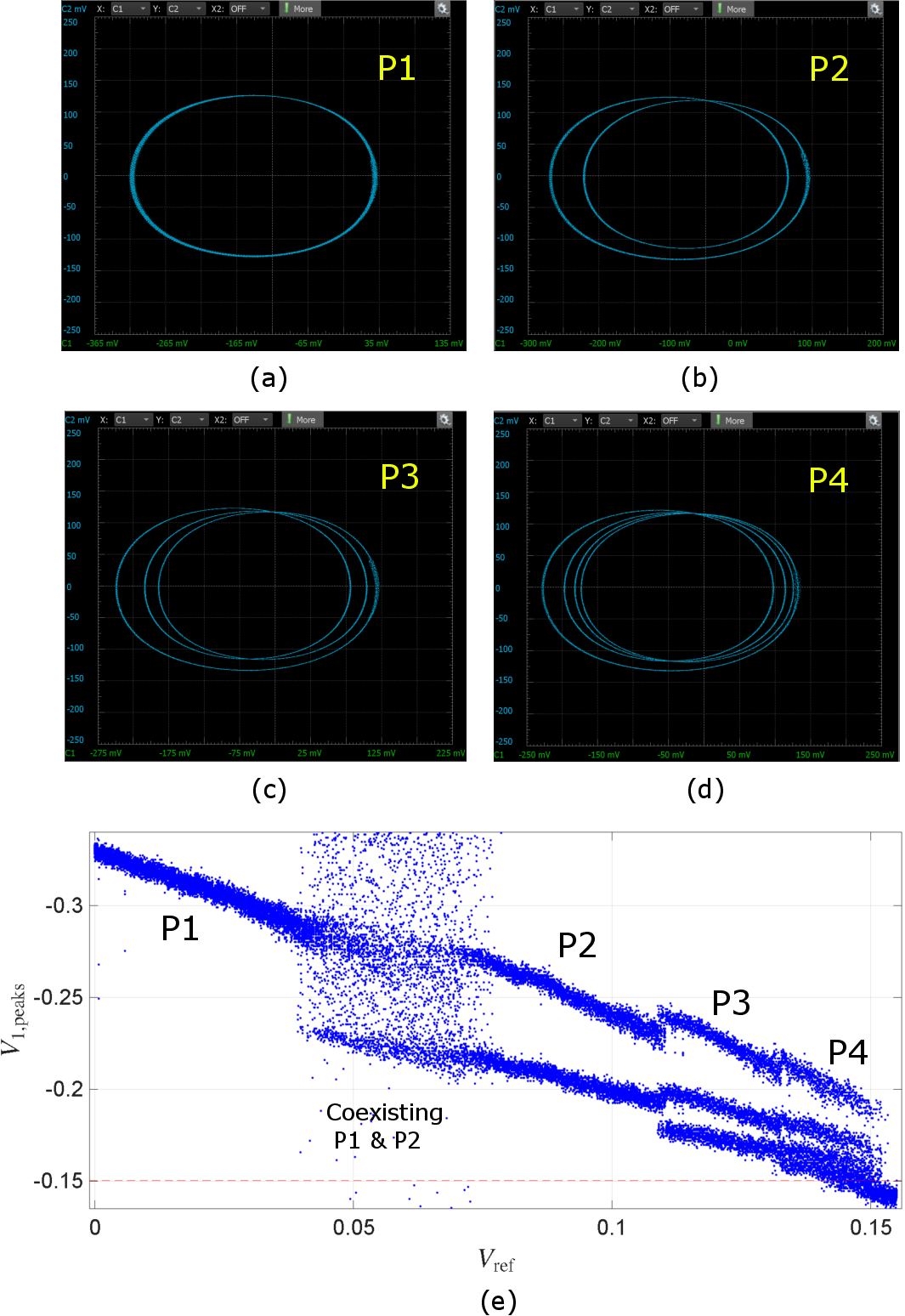}
    \caption{Experimentally obtained phase-portraits corresponding to (a)P1 for $V_{\mathrm{ref}} = 30$mV, (b) P2 for $V_{\mathrm{ref}} = 90$mV, (c) P3 for $V_{\mathrm{ref}} = 115$mV and (d) P4 for $V_{\mathrm{ref}} = 135$mV. The experimentally obtained bifurcation diagram demonstrating period-adding cascades as the barrier distance or $V_{\mathrm{ref}}$ is varied is shown in (e).}
    \label{fig 12}
\end{figure}

\noindent Figure \ref{fig 12} presents the experimentally obtained phase portraits and bifurcation diagram, demonstrating the occurrence of a period-adding cascade. The phase portraits in Figs. \ref{fig 12}(a)–(d) show P1 to P4 orbits corresponding to reference voltages $V_{\mathrm{ref}} = 30$, $90$, $115$, and $135$ mV, respectively, as presented in Fig. \ref{fig 11}. The bifurcation diagram in Fig. \ref{fig 12}(e) shows the experimentally obtained negative peaks $V_{1,\mathrm{peaks}}^y$ as the barrier location, controlled through $V_{\mathrm{ref}}$, is varied over $0.001 \leq V_{\mathrm{ref}} \leq 0.155$ V. The experimental bifurcation diagram shows close quantitative agreement with the numerical results in Fig. \ref{fig 2}, with both the response amplitudes and the parameter locations of the period-adding transitions exhibiting good correspondence, thereby validating the occurrence of a period-adding cascade due to soft impacts. The scattered experimental responses observed between the P1 and P2 branches in Fig. \ref{fig 12}(e), around $V_{\textrm{ref}} \approx 50$mV, occur within the parameter region where two stable P1 and P2 attractors coexist, as predicted numerically in Fig. \ref{fig 2}(g) and basins of attraction in Fig. \ref{fig 4}. These scattered responses may be attributed to noise-induced transitions between the coexisting attractors during experimental measurements. For each reference voltage, a MATLAB script was developed to interface with the Digilent DAQ using MATLAB's Data Acquisition Toolbox. The script automatically varied the reference voltage $V_{\mathrm{ref}}$ through the Digilent waveform generator in a loop, enabling an automated experimental parameter sweep. At each reference voltage, the time histories of $V_1^y$ and $V_2^y$ were recorded over $0.1$ s. The negative peaks of $V_1^y$ were then extracted at the zero crossings of $V_2^y$ from negative to positive values, corresponding to positive acceleration. The extracted peaks were subsequently used to construct the experimental bifurcation diagram.

\noindent The following section investigates the route to instability of the observed periodic responses and transitions to aperiodic and chaotic dynamics using Floquet multipliers and Lyapunov exponents, respectively.  

\section{Stability analysis}\label{sec 5}

A stability analysis is conducted to investigate the route to instability of the period-adding orbits observed numerically in Sec. \ref{sec 3}. The stability of periodic orbits is assessed through Floquet multiplier analysis, while the largest Lyapunov exponent is used to characterize the aperiodic orbits. Together, these measures provide a comprehensive description of the period-adding cascade and the apparently chaotic regimes interspersed between the periodic orbits. Hence, the stability analysis will give insights on how and when the underlying dynamics become unstable or chaotic as $\sigma$ and $U_r$ is varied in the parameter space.

\noindent However, a stability analysis based on the variational equation is not straightforward for the present PWS dynamical system. This difficulty arises because, when the nominal orbit impacts the barrier, the perturbation dynamics in the neighbourhood of the discontinuity boundary cannot be described solely by the Jacobian matrix governing the variational equation. Nearby perturbed trajectories generally reach the discontinuity boundary at different times and may therefore lie in different regions of the PWS phase space. The perturbations must consequently be mapped appropriately across the discontinuity so that their subsequent evolution is governed by the correct variational equation for the corresponding region, as explained below.

\begin{figure}[h!]
    \centering
    \includegraphics[width=0.8\linewidth]{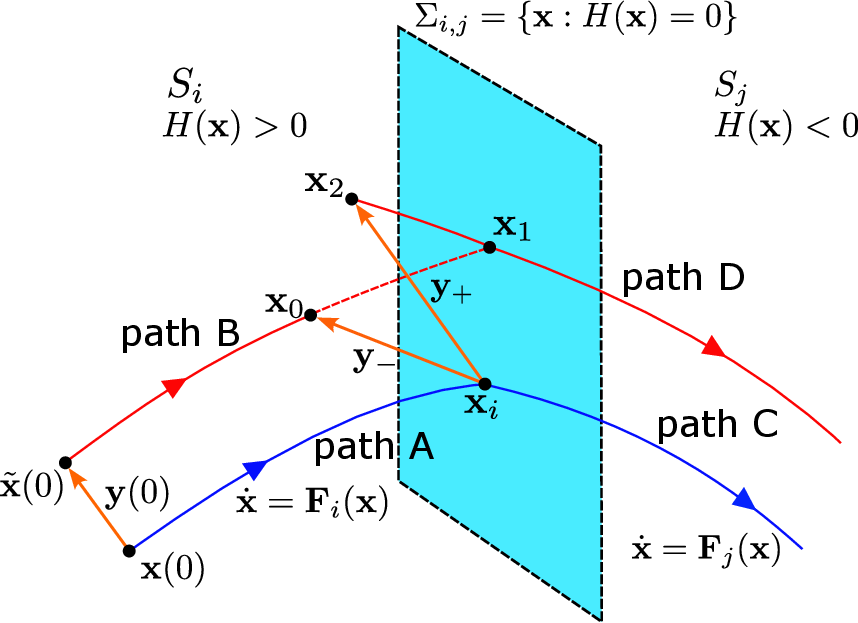}
    \caption{Schematic showing a periodic orbit $\mathbf{x}$ and a perturbed orbit $\tilde{\mathbf{x}} = \mathbf{x} + \mathbf{y}$ reaching the discontinuity boundary $\Sigma_{i,j}$ at different times. A ZTDM $\mathbf{y}_- \rightarrow \mathbf{y}_+$ is applied during an impact to ensures that perturbations $\mathbf{y}$ reach $\Sigma_{i,j}$ correctly.}
    \label{fig 13}
\end{figure}

\noindent Consider two orbits $\mathbf{x}$ and $\mathbf{x} + \mathbf{y}$, as shown in Fig. \ref{fig 13} \cite{chawla2025higher}, representing a periodic and a slightly perturbed orbit with perturbation $\mathbf{y}$. Their evolution is governed by the following equations,
\begin{align}\label{eq 5}
    \dot{\mathbf{x}} = \begin{cases}
        \mathbf{F}_i(\mathbf{x}, t) \qquad \mathbf{x} \in S_i\\
        \mathbf{F}_j(\mathbf{x}, t) \qquad \mathbf{x} \in S_j
    \end{cases}\\ 
    \dot{\mathbf{y}} = \begin{cases}
        \nabla\mathbf{F}_i^T(\mathbf{x}, t) \cdot \mathbf{y} \qquad \mathbf{y} \in S_i\\
        \nabla\mathbf{F}_j^T(\mathbf{x}, t) \cdot \mathbf{y} \qquad \mathbf{y} \in S_j
    \end{cases} \nonumber
\end{align}
where $S_i$ and $S_j$ are the subspaces defined as $S_i = \{ \mathbf{x} \in \mathbb{R}^n : H(\mathbf{x}) > 0\}$ and $S_j = \{ \mathbf{x} \in \mathbb{R}^n : H(\mathbf{x}) < 0\}$ while the discontinuity barrier is $\Sigma_{i,j} = \{ \mathbf{x} \in \mathbb{R}^n : H(\mathbf{x}) = 0\}$. Let the primary orbit $\mathbf{x} \in S_i$ impact the discontinuity boundary at $\mathbf{x}_i \in \Sigma_{i,j}$ at time $t_i$. At this instance, a perturbation $\mathbf{y} \in S_i$ reaches $\mathbf{y}_- \in S_i$. As time evolves beyond $t_i$, the nominal orbit's evolution is governed by $\mathbf{F}_j(\mathbf{x}, t)$ since $\mathbf{x} \in S_j$ after impact. However, the perturbation $\mathbf{y}_-$ has yet to reach $\Sigma_{i,j}$ and hence its evolution cannot be coupled to the Jacobian of $\mathbf{F}_j(\mathbf{x}, t)$ yet. To resolve this mismatch, a ZTDM \cite{chawla2025higher} is implemented during an impact. Let the time required for $\mathbf{y}_- \in S_i$ to reach the barrier $\Sigma_{i,j}$ at $\mathbf{x}_1 \in \Sigma_{i,j}$ be $\delta_i$. A Taylor series expansion about $\mathbf{x}_1$ back in time $-\delta_i$, assuming $\mathbf{x}_1$ is governed by $\mathbf{F}_j(\mathbf{x}, t)$, results in $\mathbf{x}_2 \in S_i$. The flight time $\delta_i$ is given by the positive root of $A\delta_i^2 + B \delta_i + C = 0$ \cite{chawla2025higher}, i.e.,
\begin{equation} \label{eq 6}
    \delta_+ = -\frac{B}{2A}\Bigg(1 - \sqrt{1 - \frac{4AC}{B^2}}\Bigg)
\end{equation}
where,
\begin{align} \label{eq 7}
	A &= \mathbf{\nabla}H(\mathbf{x}_i)^T\cdot\mathbf{\nabla}\mathbf{F_i}(\mathbf{x}_i)^T\cdot\mathbf{F_i}(\mathbf{x}_i) + \mathbf{F_i}(\mathbf{x}_i)^T\cdot\mathbf{\nabla}(\mathbf{\nabla}H(\mathbf{x}_i))^T\cdot\mathbf{F_i}(\mathbf{x}_i),\\
	B &= 2\mathbf{\nabla} H(\mathbf{x}_i)^T\cdot\mathbf{F_i}(\mathbf{x}_i) + 2\mathbf{\nabla}H(\mathbf{x}_i)^T\cdot\mathbf{\nabla}\mathbf{F_i}(\mathbf{x}_i)^T\cdot\mathbf{y}_- \nonumber \\
	&\quad + \mathbf{y}_-^T\cdot\mathbf{\nabla}(\mathbf{\nabla}H(\mathbf{x}_i))^T\cdot\mathbf{F_i}(\mathbf{x}_i) + \mathbf{F_i}(\mathbf{x}_i)^T\cdot\mathbf{\nabla}(\mathbf{\nabla}H(\mathbf{x}_i))^T\cdot\mathbf{y}_-, \nonumber \\
	C &= \mathbf{y}_-^T\cdot\mathbf{\nabla}(\mathbf{\nabla}H(\mathbf{x}_i))^T\cdot\mathbf{y}_- + 2 \mathbf{\nabla}H(\mathbf{x}_i)^T\cdot \mathbf{y}_-. \nonumber
\end{align}
\noindent The ZTDM maps the pre-perturbation state $\mathbf{y}_-$ to $\mathbf{y}_+$ which represents the post-perturbation state between $\mathbf{x}_i$ to $\mathbf{x}_2$ and is defined as \cite{chawla2025higher},
\begin{align} \label{eq 8}
	\mathbf{y}_+ = \mathbf{y}_- &+ \delta_+ \Big( \mathbf{F}_i(\mathbf{x}_i) - \mathbf{F}_j(\mathbf{x}_i) \Big) + \delta_+ \Big( \nabla \mathbf{F}_i^T(\mathbf{x}_i) - \nabla \mathbf{F}_j^T(\mathbf{x}_i) \Big) \cdot \mathbf{y}_- \nonumber \\
	&\quad + \frac{\delta_+^2}{2} \Big( \nabla \mathbf{F}_i^T(\mathbf{x}_i) \cdot \mathbf{F}_i(\mathbf{x}_i) - 2 \nabla \mathbf{F}_j^T(\mathbf{x}_i) \cdot \mathbf{F}_i(\mathbf{x}_i) + \nabla \mathbf{F}_j^T(\mathbf{x}_i) \cdot \mathbf{F}_j(\mathbf{x}_i) \Big) \nonumber \\
	&\qquad + \mathcal{O}(\delta_+^3, \delta_+^2 ||\mathbf{y}_-||, \delta_+ ||\mathbf{y}_-||^2).
\end{align}
\noindent Hence, after a ZTDM, the numerical integration of $\mathbf{x}_i$ and $y_+$ can continue using the post-impact dynamics governed by the Jacobian $\nabla \mathbf{F}_j(\mathbf{x}, t)$. The ZTDM ensures that $\mathbf{y}_+$ or $\mathbf{x}_2$ reaches $\mathbf{x}_1 \in \Sigma_{i,j}$ after a time lapse of $t_i + \delta_i$ as if $\mathbf{y}_-$ evolves to $\mathbf{x}_1$ under the pre-impact dynamics governed by the Jacobian $\nabla \mathbf{F}_i(\mathbf{x}, t)$. 

\noindent Hence, the variational equation, together with the ZTDM, can accurately determine the perturbation dynamics in the local neighbourhood of the discontinuity boundary. The subsequent sections discuss how the resulting dynamics and ZTDM can be used to determine Floquet multipliers and Lyapunov exponents for stability analysis.

\subsection{Floquet multipliers}\label{sec 5a}
This section investigates the route to instability of the observed periodic orbits by examining the corresponding Floquet multipliers. In general, for a smooth dynamical system, the eigenvalues of the corresponding monodromy matrix can determine when a periodic orbit becomes unstable as they cross the unit circle in the complex plane. Since the present dynamical system is PWS, the monodromy matrix must be constructed as a product of state transition matrices (STMs) and saltation matrices arranged according to the sequence of impact occurrences. Since Floquet theory relies on a linearized evolution of perturbations and requires a linearized mapping across an impact event, the transformation between post- and pre-impact perturbations is given by $\mathbf{y}_+ = \mathbf{S} \cdot \mathbf{y}_-$ (see Eq. \ref{eq 8}), where $\mathbf{S}$ is the saltation matrix \cite{chawla2025higher} implemented as described below.
\begin{figure}[h!]
    \centering
    \includegraphics[width=0.8\linewidth]{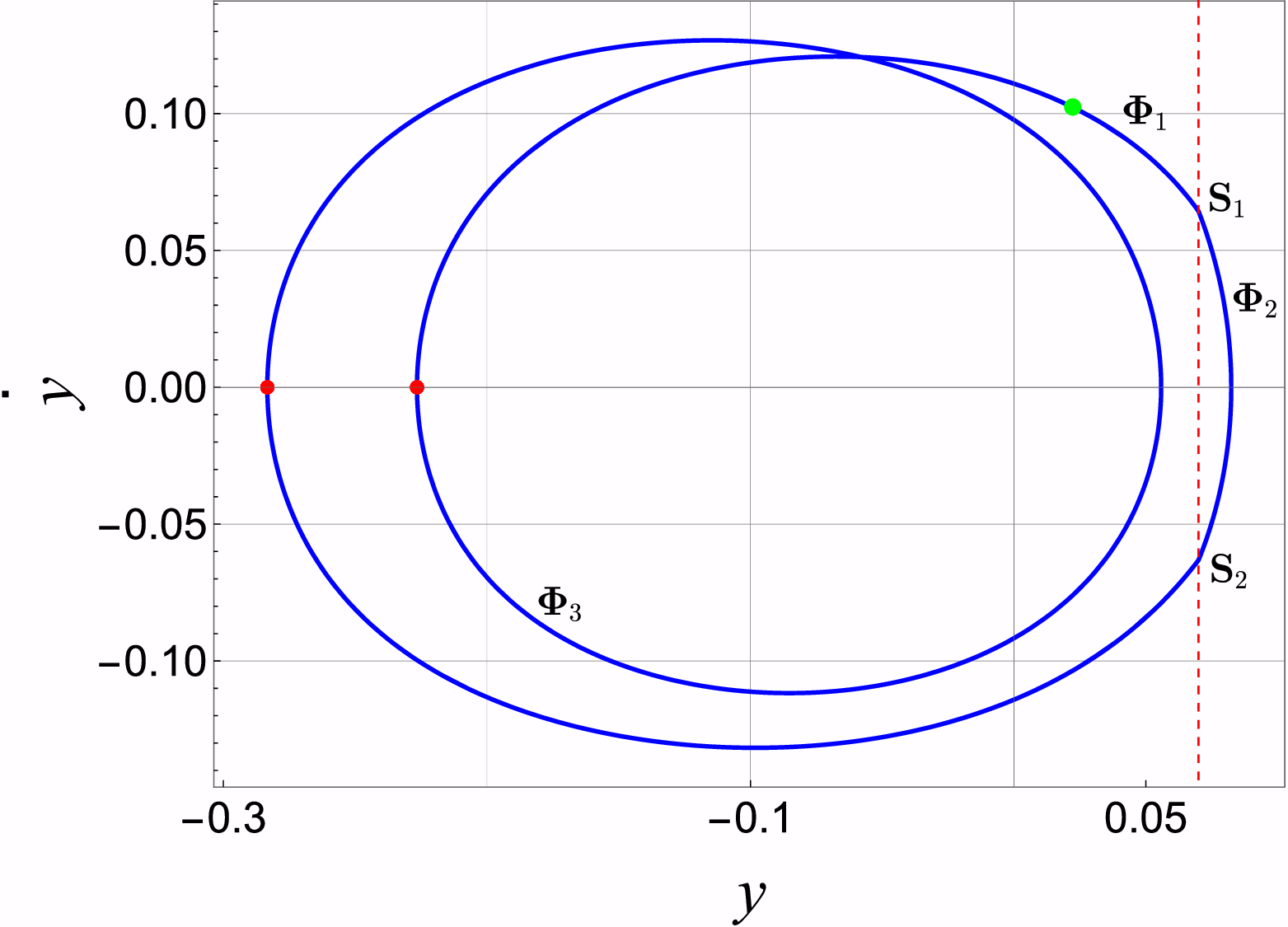}
    \caption{Schematic showing a P2 orbit for $\sigma = 0.07$, $U_r = 25$, $\beta = 25$ and $\xi_2 = 0.01$. The orbit is initiated from the green point and returns after a time period of $2T$.}
    \label{fig 14}
\end{figure}

\noindent Consider a P2 orbit shown in Fig. \ref{fig 14} corresponding to $\sigma = 0.07$, $U_r = 25.0$ and $\beta = 25$. The evolution of this periodic orbit with time period $2T$ can be decomposed into a sequence of STMs $\mathbf{\Phi}_i$ initiated from a suitable initial Poincar\'e section $\Pi$. Let $\mathbf{\Phi}_1$, $\mathbf{\Phi}_3$ and $\mathbf{\Phi}_2$ be the corresponding STMs in the subspace $S_i$ and $S_j$, away from the discontinuity boundary, respectively. The STMs of perturbations during an impact are defined using the saltation matrices $\mathbf{S}_1$ and $\mathbf{S}_2$, respectively. Then, the monodromy matrix for this P2 orbit, shown in Fig. \ref{fig 14}, is given by,
\begin{equation}\label{eq 9}
    \mathbf{\Phi}(2T) = \mathbf{\Phi}_3 \cdot \mathbf{S}_2 \cdot \mathbf{\Phi}_2 \cdot \mathbf{S}_1 \cdot \mathbf{\Phi}_1.
\end{equation}
Note that the STMs are numerically calculated using the relation,
\begin{equation}\label{eq 10}
    \dot{\mathbf{Y}} = \begin{cases}
    \nabla \mathbf{F}_i^T \cdot \mathbf{Y} \qquad \mathbf{x} \in S_i\\
    \nabla \mathbf{F}_j^T \cdot \mathbf{Y} \qquad \mathbf{x} \in S_j
    \end{cases}
\end{equation}
where $\mathbf{Y} \in \mathbb{R}^{n \times n}$ is a matrix constructed such that each column entry is a perturbation initiated from an orthonormal $n$-dimensional basis such that $\mathbf{Y}(0) = \mathbb{I}_{n \times n}$. This ensures that the fundamental solution matrix satisfies $\mathbf{\Phi}(0) = \mathbb{I}_{n \times n}$ and its value corresponds to the monodromy matrix after one time period. Therefore, for an arbitrary period $m$ orbit with orbital period $T$, the monodromy matrix is given by,
\begin{equation}\label{eq 11}
    \mathbf{\Phi}(T) = \mathbf{\Phi}_{m + 1} \prod_{i=1}^{m} \mathbf{S}_{i}\cdot \mathbf{\Phi}_i
\end{equation}
when initiated away from the discontinuity boundary. Note that the matrices are multiplied in chronological order from right to left since matrix multiplication is non-commutative. The eigenvalues of $\mathbf{\Phi}(T)$, known as Floquet multipliers, determine the orbital divergence of periodic orbits and can be used to find the route to instability. Note that the present dynamical system is autonomous due to the wake variable modeled as a Van der Pol oscillator; hence, the time period is obtained numerically as the orbit returns to a suitable Poincar\'e section. For the soft-impacting VIV system of Eq. \ref{eq 2}, the respective velocity fields are
\begin{align}
\mathbf{F}_i &= \begin{bmatrix}\label{eq 12}
                \dot{y}\\
                -\delta^2 y - (2\xi_1\delta + \frac{\gamma}{\mu})\dot{y} + Mq\\
                \dot{q}\\
                -q - \epsilon(q^2 - 1)\dot{q} + A(-\delta^2 y - (2\xi_1\delta + \frac{\gamma}{\mu})\dot{y} + Mq)
                \end{bmatrix},
                \qquad y < \sigma\\
\mathbf{F}_j &= \begin{bmatrix}
                \dot{y}\\
                -\delta^2 y - (2\xi_1\delta + \frac{\gamma}{\mu})\dot{y} + Mq - \beta \delta^2 y - 2\xi_2\dot{y}\\
                \dot{q}\\
                -q - \epsilon(q^2 - 1)\dot{q} + A(-\delta^2 y - (2\xi_1\delta + \frac{\gamma}{\mu})\dot{y} + Mq - \beta \delta^2 y - 2\xi_2\dot{y})
                \end{bmatrix},
                \qquad y \geq \sigma \nonumber
\end{align}
and the saltation matrices are
\begin{align}
\mathbf{S}_i &= \begin{bmatrix}\label{eq 13}
                \quad \quad 1\quad \quad \quad 0 \quad 0 \quad 0\\
                -\frac{(y \beta \delta^2 + 2\xi_2\dot{y})}{\dot{y}} \quad 1 \quad 0 \quad 0\\
                \qquad \quad 0\quad \quad \quad 0 \quad 1 \quad 0\\
                -\frac{A(y \beta \delta^2 + 2\xi_2\dot{y})}{\dot{y}} \quad 0 \quad 0 \quad 1
                \end{bmatrix},
                \qquad y < \sigma\\
\mathbf{S}_j &= \begin{bmatrix}
                \quad \quad 1\quad \quad \quad 0 \quad 0 \quad 0\\
                \frac{(y \beta \delta^2 + 2\xi_2\dot{y})}{\dot{y}} \quad 1 \quad 0 \quad 0\\
                \quad \quad 0\quad \quad \quad 0 \quad 1 \quad 0\\
                \frac{A(y \beta \delta^2 + 2\xi_2\dot{y})}{\dot{y}} \quad 0 \quad 0 \quad 1
                \end{bmatrix},
                \qquad y \geq \sigma \nonumber
\end{align}
corresponding to whether the trajectory approaches $\Sigma_{i,j}$ from $S_i$ or $S_j$. The algorithm for obtaining Floquet multipliers is given in Appendix \ref{app b}. 

\noindent Figures \ref{fig 15}(a) and (b) compare the real part of Floquet multipliers with the corresponding forward and backward bifurcation diagrams as the barrier distance is varied between $0.001 \leq \sigma \leq 0.150298$. The results correspond to $\beta = 25.0$, $U_r = 25.0$ and $\xi_2 = 0.01$. As the Floquet multiplier $\rho_2$ approaches and crosses the unit circle from $-1$, the corresponding periodic orbit in the bifurcation diagram becomes unstable, indicating a period-doubling bifurcation. When following the P1 orbit in the forward bifurcation diagram of Fig. \ref{fig 15}(a), in the parameter region where P1 and P2 orbits coexist, $\rho_2$ crosses $-1$ at $\sigma = 0.04743$. At this point, the P1 orbit becomes unstable, and a P2 orbit appears immediately. The Floquet multipliers jump to values with norm less than $1$ and increase with $\sigma$. Similarly, in the backward bifurcation diagram of Fig. \ref{fig 15}(b), in the coexisting P1--P2 region, $\rho_2$ corresponding to P2 orbits jumps discontinuously at $\sigma = 0.0391$ within $\pm 1$ when the P2 orbit disappears, since the P1 orbit coexists in the bifurcation diagram. Therefore, for all numerical results shown next, the forward bifurcations with increasing parameter will be considered.

\noindent Note that the value of one Floquet multiplier, in all the numerical results shown, remains unity, consistent with the Floquet theory for autonomous dynamical systems. The perturbation tangent to the periodic orbit corresponds to an infinitesimal time shift along the same trajectory, and therefore neither grows nor decays over one complete period. The tangent direction is an eigenvector of the monodromy matrix associated with the Floquet multiplier $\rho = 1$. Hence, the results confirm that incorporating the ZTDM during impact crossing yields the true monodromy matrix.

\begin{figure}
    \centering
    \includegraphics[width=1.0\linewidth]{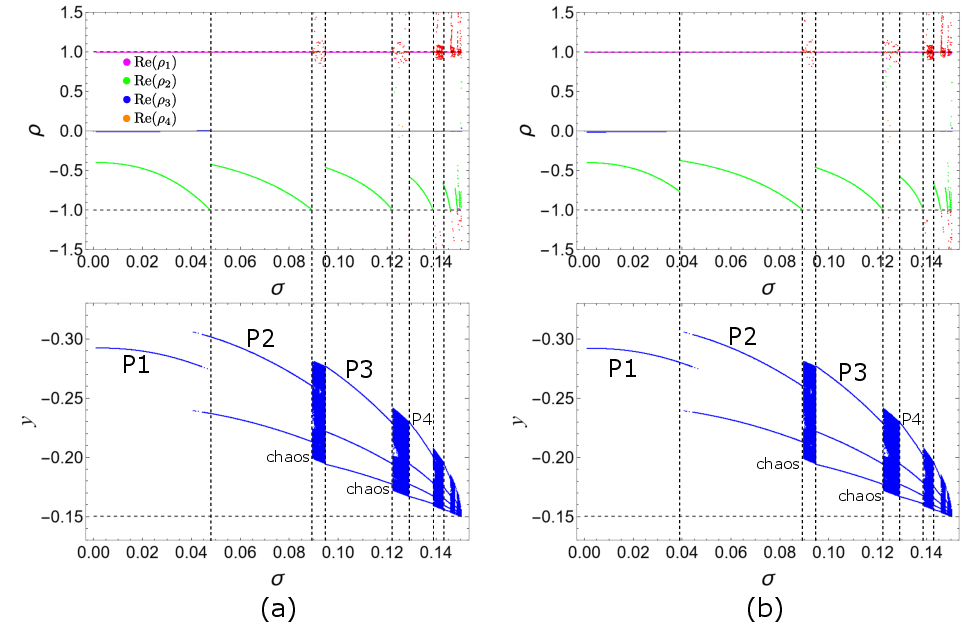}
    \caption{Comparison of Floquet multipliers $\rho$ (top) vs. barrier distance $\sigma$ with the corresponding (a) forward and (b) backward bifurcation diagrams (bottom). The results correspond to $\beta = 25$, $U_r = 25.0$ and $\xi_2 = 0.01$.}
    \label{fig 15}
\end{figure}

\begin{figure}
    \centering
    \includegraphics[width=1.0\linewidth]{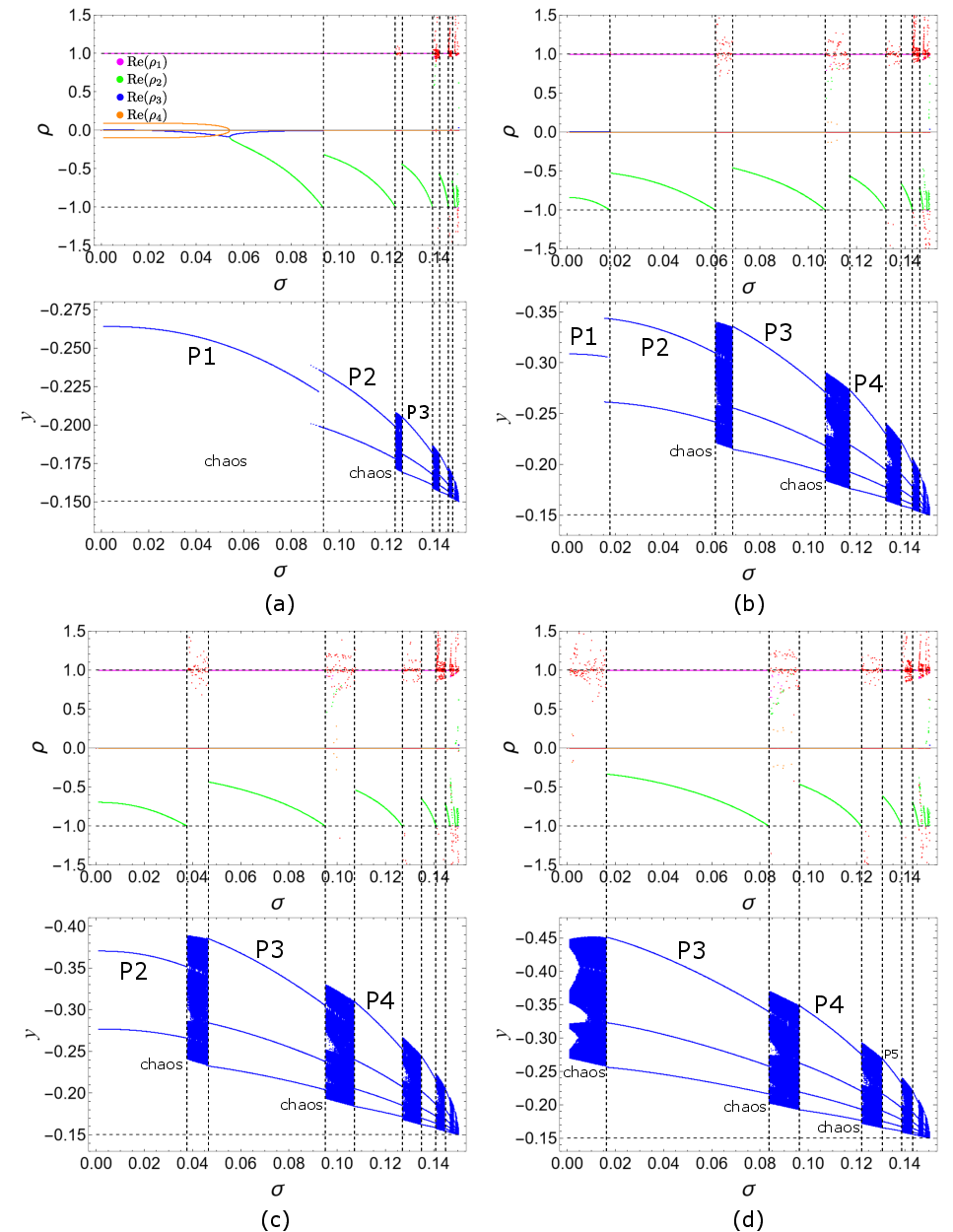}
    \caption{Comparison of Floquet multipliers $\rho$ (top) vs. barrier distance $\sigma$ with the forward bifurcation diagrams (bottom) for the stiffness ratios (a)$\beta = 10$, (b)$\beta = 50$, (c)$\beta = 100$, and (d)$\beta = 500$. The results correspond to $U_r = 25.0$ and $\xi_2 = 0.01$.}
    \label{fig 16}
\end{figure}

\noindent Figures \ref{fig 16}(a)-(d) compare the Floquet multipliers with the respective forward bifurcation diagrams as the barrier distance is varied for different stiffness ratios with $U_r = 25.0$ and $\xi_2 = 0.01$. The results shown in Figs. \ref{fig 16}(a)-(d) correspond to $\beta = 10$, $50$, $100$ and $500$ respectively. For each stiffness ratio, the numerically observed periodic orbits become unstable when the corresponding Floquet multiplier $\rho_2$ crosses $-1$, indicating a period-doubling bifurcation. The results remain consistent as the stiffness ratio increases, including near the grazing region, where higher-period orbits appear as $\sigma$ approaches $0.150298$ and likewise become unstable via period-doubling.

\begin{figure}
    \centering
    \includegraphics[width=1.0\linewidth]{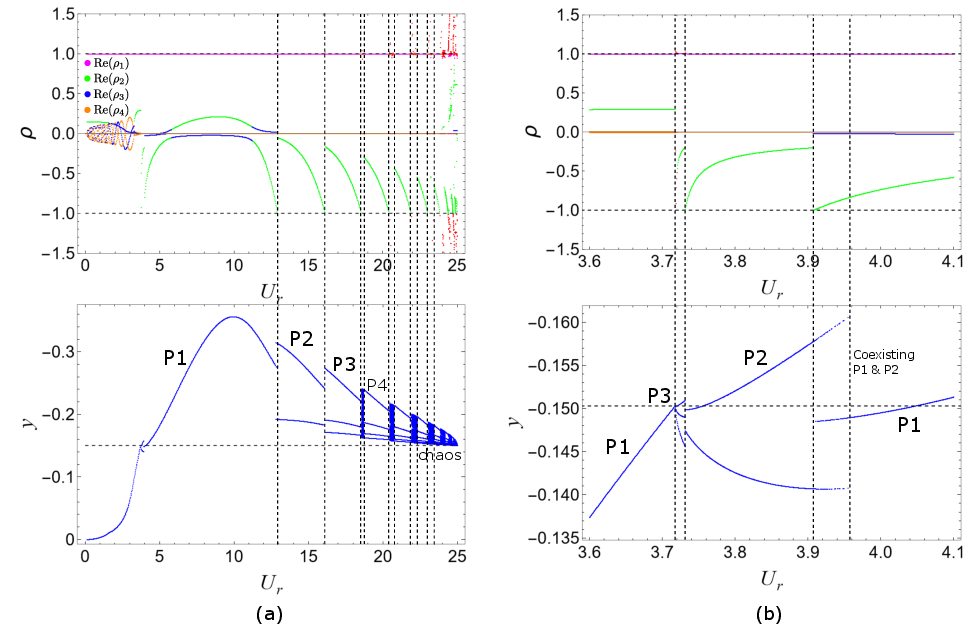}
    \caption{Comparison of Floquet multipliers $\rho$ (top) vs. reduced velocity $U_r$ with the forward bifurcation diagram (bottom) when $U_r$ varies between (a) $0.1 \leq U_r \leq 25.0$ and (b) $3.6 \leq U_r \leq 4.1$. The results correspond to $\beta = 500$, $\sigma = 0.150298$ and $\xi_2 = 0.01$.}
    \label{fig 17}
\end{figure}

\noindent Figures \ref{fig 17}(a) and (b) show the Floquet multipliers as the reduced velocity $U_r$ is varied with the barrier kept fixed at $\sigma = 0.150298$, stiffness ratio of $\beta = 500$, and $\xi_2 = 0.01$. Figs. \ref{fig 17}(a) and (b) correspond to the upper and lower branches, varying between $0.1 \leq U_r \leq 25.0$, and the initial--upper branch, varying between $3.6 \leq U_r \leq 4.1$. As the reduced velocity varies, the periodic orbits become unstable via a period-doubling bifurcation as the Floquet multiplier $\rho_2$ crosses $-1$. The results are consistent for all values of $U_r$ investigated, including the near-grazing region as $U_r$ approaches $25.0$. Near the initial--upper branch shown in Fig. \ref{fig 17}(b), where a reverse period-adding sequence appears, the P1 orbit becomes unstable via period-doubling and a P2 orbit is obtained immediately. As $U_r$ decreases further, the P2 orbit becomes unstable, since $\rho_2$ crosses $-1$, and a P3 orbit is subsequently observed.

\noindent The following section investigates the aperiodic orbits that emerge immediately after a period-doubling bifurcation by estimating their largest Lyapunov exponents.

\subsection{Lyapunov exponents}\label{sec 5b}
The largest Lyapunov exponents are estimated to complement the Floquet analysis and further characterize the stability of the present dynamical system. Since Floquet theory is applicable only to periodic orbits, the largest Lyapunov exponent is used to characterize the aperiodic orbits observed immediately after a period-doubling bifurcation, where a Floquet multiplier crosses the unit circle through $-1$. 
For the present PWS dynamical system, across an impact event, the perturbations are mapped from pre-impact state $\mathbf{y}_-$ to post-impact state $\mathbf{y}_+$ using the saltation matrix \cite{chawla2024discontinuity,chawla2025higherpoincare,chawla2025higher,chawla2023higher} defined in Eq. \ref{eq 13}.

\noindent To estimate Lyapunov exponents, perturbations are initialized along an orthonormal $n$-dimensional basis with initial norm $r_0 = 10^{-6}$ in the local neighborhood of the dynamical state $\mathbf{x}_0$. As the perturbations evolve, the initial hypersphere deforms into a hyper-ellipsoid over a stroboscopic time interval $\tau$. After each stroboscopic interval $\tau$, a QR decomposition (QRD) of the perturbation matrix $\mathbf{Y} = \mathbf{Q}\mathbf{R}$ is performed, where $\mathbf{Q}$ is an orthogonal matrix and $\mathbf{R}$ is an upper-triangular matrix. The QRD implements Gram-Schmidt orthogonalization, ensuring that the perturbations remain mutually orthogonal. The stretching factors $r_i = |R_{ii}|$ for each orthogonalized perturbation direction are obtained from the absolute values of the diagonal entries of $\mathbf{R}$. The perturbations are then reinitialized with norm $r_0$ while retaining the orthogonalized directions, and the process is repeated. The Lyapunov exponents $\lambda_i$ can now be estimated using,
\begin{equation}\label{eq 14}
    \lambda_i = \frac{1}{k\tau}\sum_{j = 1}^k \ln{\frac{r^{(j)}_i}{r_0}}
\end{equation}
where $r_i^{(j)} = |R_{ii}^{(j)}|$ is the stretching factor associated with the $i$th orthogonalized perturbation direction, obtained from the absolute value of the corresponding diagonal entry of $\mathbf{R}$ at the $j$th stroboscopic interval. The algorithm to estimate the Lyapunov exponents is given in Appendix \ref{app c}. 

\noindent Figure \ref{fig 18} compares the Lyapunov spectrum with the corresponding bifurcation diagram as the barrier distance is varied between $0.001 \leq \sigma \leq 0.150298$ with $U_r = 25.0$, $\beta = 25.0$ and $\xi_2 = 0.01$. In Fig. \ref{fig 18}(a), the Lyapunov spectrum is shown as the barrier distance is varied. The exponents have been normalized in \ref{fig 18}(a) for an improved visual representation. In Fig. \ref{fig 18}(b), the largest Lyapunov exponent and the second Lyapunov exponent are shown as $\sigma$ is varied. The largest Lyapunov exponent always remains zero for periodic orbits since the PWS soft-impacting system is autonomous. The largest Lyapunov exponent becomes positive within the aperiodic windows in the respective bifurcation diagram, confirming that the orbits are chaotic. The second Lyapunov exponent remains negative for periodic orbits and approaches zero as the orbits become unstable through a period-doubling bifurcation, as indicated by the respective Floquet analysis.

\begin{figure}
    \centering
    \includegraphics[width=1.0\linewidth]{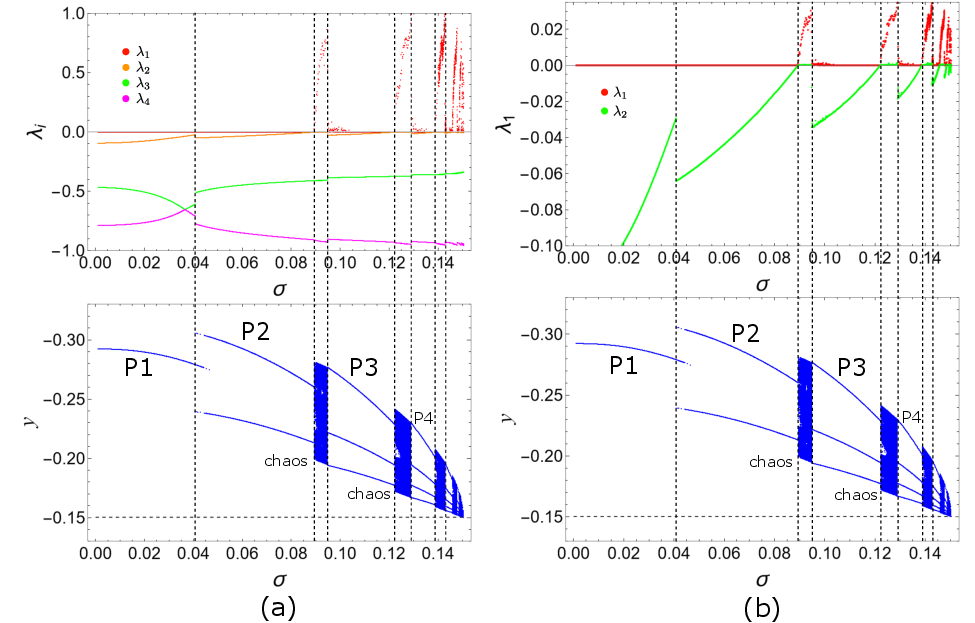}
    \caption{Comparison of (a) normalized Lyapunov exponents $\lambda_i$ (top) and (b) largest Lyapunov exponents $\lambda_1$ and $\lambda_2$ vs. barrier distance $\sigma$ with the corresponding forward bifurcation diagrams (bottom). The results correspond to $\beta = 25$, $U_r = 25.0$ and $\xi_2 = 0.01$.}
    \label{fig 18}
\end{figure}

\begin{figure}
    \centering
    \includegraphics[width=1.0\linewidth]{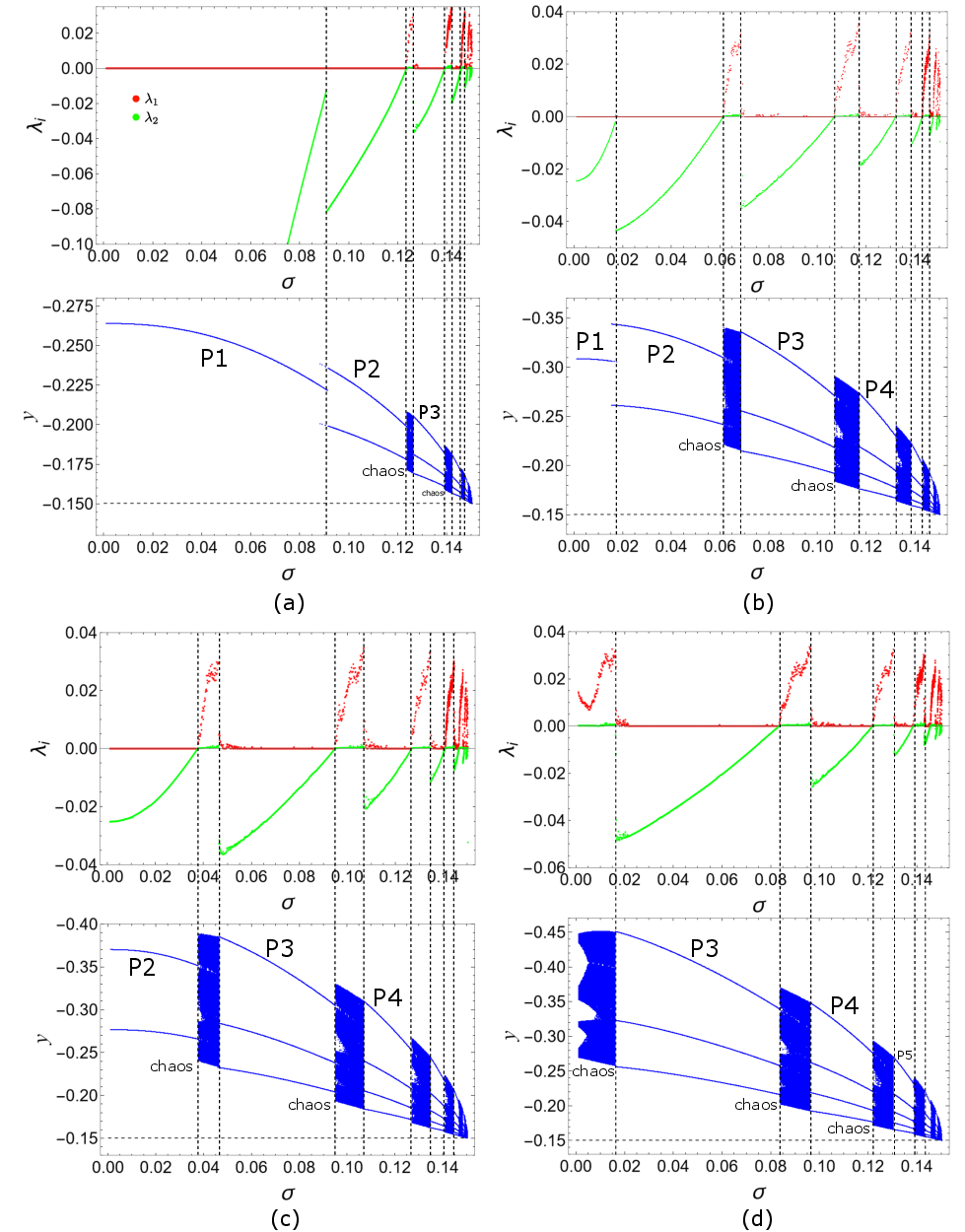}
    \caption{Comparison of the largest Lyapunov exponent $\lambda_1$ and $\lambda_2$ (top) vs. barrier distance $\sigma$ with the corresponding forward bifurcation diagram (bottom) for the stiffness ratios (a)$\beta = 10$, (b)$\beta = 50$, (c)$\beta = 100$, and (d)$\beta = 500$. The results correspond to $U_r = 25.0$ and $\xi_2 = 0.01$.}
    \label{fig 19}
\end{figure}

\noindent Similarly, Fig. \ref{fig 19}(a) - (d) shows the largest and second Lyapunov exponents as the barrier distance is varied for $\beta = 10$, $50$, $100$ and $500$, respectively, with $U_r = 25.0$ and $\xi_2 = 0.01$. The largest Lyapunov exponent remains zero for periodic orbits and becomes positive between the observed period-adding cascade of orbits. The second Lyapunov exponent approaches zero, indicating the route to instability, as confirmed by the corresponding bifurcation diagram. The ZTDM, used to estimate Lyapunov exponents, correctly captures the perturbation dynamics in the local neighborhood of the discontinuity boundary. The results are consistent with the respective bifurcation diagrams over a wide range of stiffness ratios between the cylinder and the deformable barrier.

\begin{figure}
    \centering
    \includegraphics[width=1.0\linewidth]{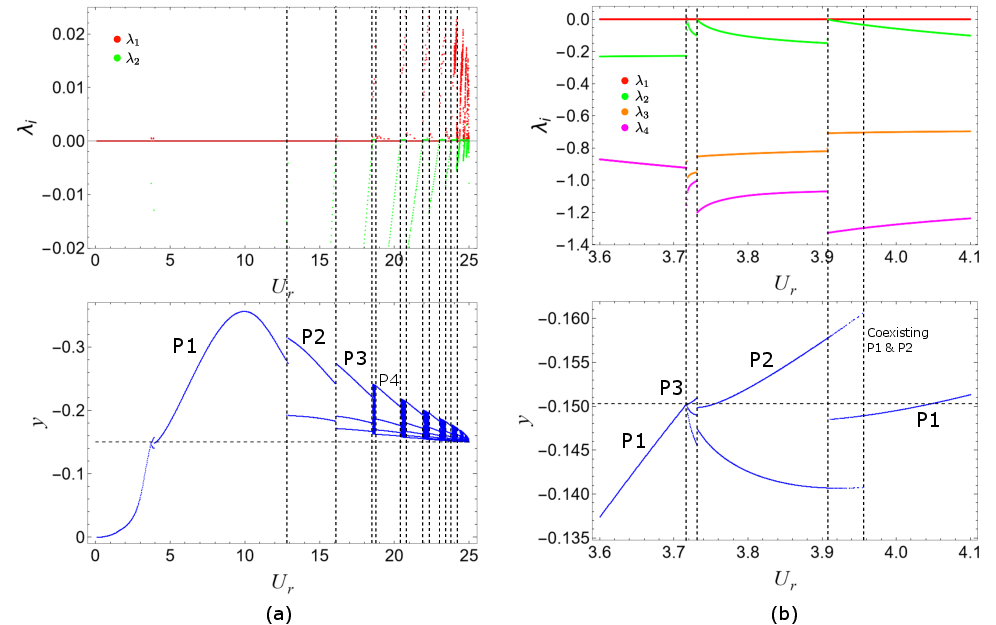}
    \caption{Comparison of the largest Lyapunov exponent $\lambda_1$ and $\lambda_2$ (top) vs. reduced velocity $U_r$ with the forward bifurcation diagram (bottom) when $U_r$ varies between (a) $0.01 \leq U_r \leq 25.0$ and (b) $3.6 \leq U_r \leq 4.1$. The results correspond to $\beta = 500$, $\sigma = 0.150298$ and $\xi_2 = 0.01$.}
    \label{fig 20}
\end{figure}

\noindent Figure \ref{fig 20} compares the Lyapunov exponents with the respective bifurcation diagrams as the reduced cross-flow velocity is varied between $0.01 \leq U_r \leq 25.0$ for $\sigma = 0.150298$, $\beta = 500$ and $\xi_2 = 0.01$. In Fig. \ref{fig 20}(a), the largest Lyapunov exponent becomes positive for the windows of $U_r$ where aperiodic orbits are seen, concluding that the orbits are chaotic. In Fig \ref{fig 20}(b), no positive Lyapunov exponents are observed, indicating that no chaotic orbits exist in the initial-upper branches ranging between $3.6 \leq U_r \leq 4.1$. As a reverse period-adding sequence, i.e., P3 - P2 - P1, is observed, the second Lyapunov exponent approaches zero when orbits become unstable and immediately jumps to a negative value as the cylinder exhibits oscillations corresponding to an existing stable orbit. This is also consistent with the Floquet analysis, where a period-doubling bifurcation occurs at the corresponding parameter values for which the second Lyapunov exponent approaches zero.

\noindent The following section summarizes the main findings of this study and presents the concluding remarks.

\section{Conclusions}\label{sec 6}
This article investigates the impact dynamics of a structure subjected to VIVs during the onset of wake. The cylindrical structure is modeled as a harmonic oscillator, while the self-sustained lift coefficient responsible for the transverse fluid forcing is represented using a Van der Pol oscillator. A soft deformable barrier is placed near the structure and contacts it at sufficiently large cross-flow response amplitudes, introducing an additional restoring force. The resulting contact dynamics is modeled using soft impacts, leading to a PWS bilinear oscillator represented as a Filippov-type dynamical system. A systematic numerical investigation varies the stiffness ratio, barrier distance, and fluid-flow velocity to characterize how soft impacts influence the system dynamics.

\noindent Periodic orbits intersecting a Poincar\'e section are identified and labeled as P1, P2, and higher-period responses. Numerical results show that variations in barrier distance and cross-flow velocity give rise to DIBs in the form of period-adding cascades, with periodic orbits interspersed with chaotic responses. As the stiffness ratio between the barrier and the structure increases, the period-adding cascade shifts towards higher-periodic responses while the lower P1 and P2 orbits disappear. Higher-period orbits also accumulate as the barrier distance approaches the grazing condition. A similar period-adding cascade and accumulation of higher-period responses occur when the fluid-flow velocity increases while the stiffness ratio remains fixed. Coexisting P1 and P2 attractors are observed across the investigated cases and are confirmed using the corresponding basins of attraction. Furthermore, when the fluid-flow velocity is varied following the onset of soft impact near the initial branch, an inverted sequence consisting of P3, P2, and P1 periodic orbits is observed.

\noindent Experimental investigations of period-adding cascades in autonomous PWS dynamical systems remain relatively limited. Therefore, the numerically observed DIBs and period-adding cascades are experimentally validated using an analog electronic circuit. An inductorless circuit comprising op-amps configured as integrators is implemented, and the complete circuit is fabricated on a printed circuit board to reduce unwanted electrical noise. Experimentally obtained voltage responses, phase portraits and period-added cascades using a digital oscilloscope show good agreement with the corresponding numerical results. In addition, an automated experimental procedure is developed using MATLAB and Digilent data-acquisition hardware to construct bifurcation diagrams under parametric variation.

\noindent The stability and route to instability of the observed periodic responses are further investigated using Floquet multipliers and Lyapunov exponents. To account for the discontinuities introduced by soft impacts, a flight-time correction together with the ZTDM is incorporated into the perturbation dynamics. The Floquet multipliers are found to cross the unit circle through $-1$, demonstrating that the periodic orbits lose stability through period-doubling bifurcations. No fold bifurcations are observed along the investigated period-adding cascades. For the aperiodic responses between successive periodic windows, the largest Lyapunov exponent remains positive, confirming chaotic dynamics.

\noindent Overall, the results demonstrate that soft impacts can produce rich DIBs in vortex-induced vibration systems, including period-adding cascades, coexistence of attractors, accumulation of higher-period responses near grazing, and chaotic windows generated through period-doubling instability. These findings provide insight into the parameter regimes in which higher-period and chaotic responses may arise in structures undergoing combined vortex-induced vibration and intermittent contact, and may aid the interpretation and design of systems in which impact constraints are introduced to limit large-amplitude VIV responses.

\bibliography{references}

\begin{appendices}

\section{Nomenclature}\label{app a}

\begin{table}[!]
\caption{Description of variable names and symbols}
\begin{tabular}{@{}lll@{}}
    \toprule
    Variable name & Symbols & Units\\
    \midrule
    \textbf{Structure variables} & &\\
    Displacement of structure & $Y$ & m\\
    Diameter of structure & $D$ & m\\
    Mass of oscillator & $m$ & kg\\
    Mass of structure & $m_s$ & kg\\
    Fluid-added mass & $m_f$ & kg\\
    Stiffness of oscillator $Y$ & h & N $m^{-1}$\\
    Damping of oscillator & $r$ & Ns $m^{-1}$\\
    Damping of structure & $r_s$ & Ns $m^{-1}$\\
    Fluid-added damping & $r_f$ & Ns $m^{-1}$\\
    Time & $\tau$ & s\\
    Lift force & $S$ & N\\
    Natural frequency of structure & $\Omega_s$ & $s^{-1}$\\
    Fluid-added mass coefficient & $C_m$ & -\\
    Fluid flow velocity & $U$ & m$s^{-1}$\\
    Density of fluid medium & $\rho$ & kg $m^{-3}$\\
    \textbf{Wake variables} & &\\
    Reduced vortex lift coefficient & $q$ & -\\
    Vortex shedding frequency & $\Omega_f$ & $s^{-1}$\\
    Nonlinear parameter & $\epsilon$ & -\\
    Coupling force of structure on wake & $F$ & N\\
    Vortex lift coefficient & $C_L$ & -\\
    Reference lift coefficient & $C_{Lo}$ & -\\
    Strouhal number & $S_t$ & -\\
    \textbf{Dimensionless variables} & &\\
    Structure displacement & $y$ & -\\
    Stopper location & $\sigma$ & -\\
    Reduced vortex lift coefficient & $q$ & -\\
    Stroboscopic displacement & $y_0$ & -\\
    Time & $t$ & -\\
    Ratios of oscillator frequency & $\delta$ & -\\
    Damping ratio & $\xi_1$ & -\\
    Stall parameter & $\gamma$ & -\\
    Mass ratio & $\mu$ & -\\
    Reduced coupling force on wake & $f$ & -\\
    Coupling parameter & $A$ & -\\
    Reduced fluid-flow velocity & $U_r$ & -\\
    Reduced lift force & $s$ & -\\
    Reduced coupling variable & $M$ & -\\
    Stiffness of soft barrier & $\beta$ & -\\
    Damping ratio of soft barrier & $\xi_2$ & -\\
    \textbf{Piecewise-smooth variables} & & \textbf{Dims.}\\
    Generalized state & $\mathbf{x}$ & $\mathbb{R}^{n}$\\
    Generalized vector field before impact& $\mathbf{F}_i(\mathbf{x})$ & $\mathbb{R}^{n}$\\
    Generalized vector field after impact& $\mathbf{F}_j(\mathbf{x})$ & $\mathbb{R}^{n}$\\
    Perturbation to $\mathbf{x}$ & $\mathbf{y}$ & $\mathbb{R}^{n}$\\
    Discontinuity constraint & $H(\mathbf{x})$ & $\mathbb{R}^{n - 1}$\\
    Discontinuity boundary & $\mathbf{\Sigma_{i,j}}$ & $\mathbb{R}^{n - 1}$\\
    Poincar\'e section & $\mathbf{\Sigma_1}$ & $\mathbb{R}^{n - 1}$\\
    State $\mathbf{x}$ at impact & $\mathbf{x}_i$ & $\mathbb{R}^{n}$\\
    Perturbed state at impact & $\mathbf{x}_0$ & $\mathbb{R}^{n}$\\
    Perturbed state on $\Sigma_{i,j}$ & $\mathbf{x}_1$ & $\mathbb{R}^{n}$\\
    Mapped perturbed state via ZTDM & $\mathbf{x}_2$ & $\mathbb{R}^{n}$\\
    Instant of impact state $\mathbf{x}_i$ & $t_i$ & $\mathbb{R}^{1}$\\
    Flight time to impact state $\mathbf{x}_1$ & $\delta_+$ & $\mathbb{R}^{1}$\\
    Perturbation during impact & $\mathbf{y}_-$ & $\mathbb{R}^{n}$\\
    Mapped perturbation via ZTDM & $\mathbf{y}_+$ & $\mathbb{R}^{n}$\\
    Saltation matrix & $\mathbf{S}_{i/j}$ & $\mathbb{R}^{n \times n}$
\end{tabular}
\end{table}

\section{Algorithm for Floquet multipliers}\label{app b}

\begin{algorithm}[H]
\setstretch{0.9}
    \caption{Floquet multipliers from monodromy matrix $\mathbf{\Phi}$}
    \label{algo 1}
    \begin{algorithmic}[1]

    \State Initialize: $\mathbf{x}(0)$
    \Comment{Set initial conditions}

    \State Initialize: $\mathbf{\Phi}(0) = \mathbb{I}_{n \times n}$
    \Comment{Initialize monodromy matrix as identity}

    \While{$t \leq t_{\textrm{transient}}$}
        \Comment{Integrate for a sufficiently long time to discard transients}

        \State Integrate
        $\dot{\mathbf{x}} =
        \begin{cases}
        \mathbf{F}_i(\mathbf{x}) & \mathbf{x} \in S_i\\
        \mathbf{F}_j(\mathbf{x}) & \mathbf{x} \in S_j
        \end{cases}$

        \If{$t \geq t_{\textrm{transient}}$ and $P(\mathbf{x}) = 0$}

            \State Store
            $\mathbf{x}_{\textrm{temp}}[k] \gets \mathbf{x}(t)$
            and $t_{\textrm{temp}}[k] \gets t$
            \Comment{Store successive Poincar\'e crossings}

            \If{$\|\mathbf{x}_{\textrm{temp}}[k] - \mathbf{x}_{\textrm{temp}}[0]\| < \varepsilon_{\textrm{tol}}$}

                \State $\mathbf{x}_p \gets
                \mathbf{x}_{\textrm{temp}}[0]$

                \State $T \gets t_{\textrm{temp}}[k] - t_{\textrm{temp}}[0]$
                \Comment{Obtain periodic state and orbital period}

                \State \textbf{Break}

            \EndIf
        \EndIf
    \EndWhile

    \State Initialize: $\mathbf{x}(t_0) \gets \mathbf{x}_p$
    \Comment{Initialize on the periodic orbit}

    \State Initialize:
    $\mathbf{\Phi} \gets \mathbb{I}_{n \times n}$,
    $\mathbf{Y}(t_0) \gets \mathbb{I}_{n \times n}$
    \Comment{Initialize monodromy and variational matrices}

    \While{$t < t_0 + T$}
        \Comment{Integrate over one complete orbital period}

        \State Integrate
        $\dot{\mathbf{x}} =
        \begin{cases}
        \mathbf{F}_i(\mathbf{x}) & \mathbf{x} \in S_i\\
        \mathbf{F}_j(\mathbf{x}) & \mathbf{x} \in S_j
        \end{cases}$
        \Comment{Integrate on the limit cycle}

        \State Integrate
        $\dot{\mathbf{Y}} = \begin{cases}
        (\nabla\mathbf{F}_i(\mathbf{x}))^T
        \cdot\mathbf{Y} & \mathbf{x} \in S_i\\
        (\nabla\mathbf{F}_j(\mathbf{x}))^T
        \cdot\mathbf{Y} & \mathbf{x} \in S_j
        \end{cases}$
        \Comment{Calculate the STMs}

        \If{$H(\mathbf{x}) = 0$}
            \Comment{Occurrence of a soft impact}

            \State $\mathbf{\Phi}_i \gets \mathbf{Y}$
            \Comment{Store STM before switching}

            \State Evaluate the corresponding
            saltation matrix $\mathbf{S}_i$ or $\mathbf{S}_j$

            \State $\mathbf{y}_+ \gets
            \mathbf{S}_{i/j}\cdot\mathbf{y}_-$
            \Comment{Map perturbations across the boundary}

            \State $\mathbf{\Phi} \gets
            \mathbf{S}_{i/j}\cdot\mathbf{\Phi}_i
            \cdot\mathbf{\Phi}$
            \Comment{Update monodromy matrix}

            \State Reinitialize:
            $\mathbf{Y} \gets \mathbb{I}_{n \times n}$
            \Comment{Reinitialize STM to identity}

        \EndIf
    \EndWhile

    \State $\mathbf{\Phi} \gets
    \mathbf{Y}\cdot\mathbf{\Phi}$
    \Comment{Include the final STM}

    \State Evaluate eigenvalues of $\mathbf{\Phi}$
    to obtain the Floquet multipliers.

    \end{algorithmic}
\end{algorithm}

\section{Algorithm for Lyapunov exponents}\label{app c}

\begin{algorithm}[H]
\setstretch{0.9}
    \caption{Lyapunov exponents incorporating the ZTDM}
    \label{algo 2}
    \begin{algorithmic}[1]

    \State Initialize: $\mathbf{x}(0) \in \mathbb{R}^{n}$
    \Comment{Set initial conditions}

    \State Initialize: $\mathbf{Y}(0) = \mathbb{I}_{n \times n}$
    \Comment{Initialize $n$ orthonormal perturbation states}

    \State $\mathbf{Y}(0) \gets r_0 \mathbf{Y}(0)$
    \Comment{Scale perturbations to radius $r_0$}

    \State Initialize: $r_i \gets \{\}$, $i=1,\ldots,n$
    \Comment{Initialize lists of stretching radii}

    \While{$t \leq t_{\textrm{stop}}$}
        \Comment{Integrate up to the maximum time}

        \State Integrate
        $\dot{\mathbf{x}} =
        \begin{cases}
        \mathbf{F}_i(\mathbf{x}) & \mathbf{x} \in S_i\\
        \mathbf{F}_j(\mathbf{x}) & \mathbf{x} \in S_j
        \end{cases}$
        \Comment{Integrate dynamical system}

        \State Integrate
        $\dot{\mathbf{Y}} =
        \begin{cases}
        (\nabla\mathbf{F}_i(\mathbf{x}))^T
        \cdot\mathbf{Y} & \mathbf{x} \in S_i\\
        (\nabla\mathbf{F}_j(\mathbf{x}))^T
        \cdot\mathbf{Y} & \mathbf{x} \in S_j
        \end{cases}$
        \Comment{Evolution of perturbation states}

        \If{$H(\mathbf{x}) = 0$}
            \Comment{Occurrence of a soft impact}

            \State Evaluate the saltation matrix $\mathbf{S}_i$ or $\mathbf{S}_j$

            \State $\mathbf{y}_{k,+} \gets
            \mathbf{S}_{i/j}\cdot\mathbf{y}_{k,-}$,
            \quad $k=1,\ldots,n$
            \Comment{Map perturbations across the discontinuity boundary}

            \State $\mathbf{Y}^{+} \gets
            \mathbf{S}_{i/j}\cdot\mathbf{Y}^{-}$
            \Comment{Update perturbation matrix}

        \EndIf

        \If{$t = kT$ and $t > t_{\textrm{transient}}$}
            \Comment{Stroboscopic reorthonormalization}

            \State $[\mathbf{Q},\mathbf{R}]
            \gets \operatorname{QRD}(\mathbf{Y})$
            \Comment{Perform QR decomposition}

            \For{$i=1,\ldots,n$}

                \State $r_i[k] \gets |R_{ii}|$
                \Comment{Store stretching radius}

            \EndFor

            \State $\mathbf{Y} \gets r_0\mathbf{Q}$
            \Comment{Reinitialize orthonormal perturbations}

        \EndIf

    \EndWhile

    \For{$i=1,\ldots,n$}

        \State $L_i[k] \gets
        \displaystyle\sum_{j=1}^{k}
        \ln\left(\frac{r_i[j]}{r_0}\right)$
        \Comment{Cumulative logarithmic stretching}

        \State $\mathrm{LE}_i[k] \gets
        \displaystyle\frac{L_i[k]}{kT}$
        \Comment{Evaluate partial-sum Lyapunov exponents}

        \State $\lambda_i \gets
        \left\langle \mathrm{LE}_i[k]\right\rangle_{\textrm{last }10\%}$
        \Comment{Average the last 10\% of estimates}

    \EndFor

    \State \Return $\lambda_1,\ldots,\lambda_n$
    \Comment{Complete Lyapunov spectrum}

    \end{algorithmic}
\end{algorithm}

\end{appendices}
\end{document}